\pdfoutput=1
\documentclass[11pt]{article}
\usepackage{epsfig}
\usepackage{graphicx}
\newcommand{\be}{\begin{equation}}
\newcommand{\ee}{\end{equation}}
\newcommand{\kp}{\mbox{K$^+$~}}
\newcommand{\km}{\mbox{K$^-$~}}
\newcommand{\Apart}[1][ ]{$A_{part}${#1}}
\newcommand{\pip}[1][ ]{$\pi^{+}${#1}}
\newcommand{\pim}[1][ ]{$\pi^{-}${#1}}
\newcommand{\AGeV}[1][ ]{$A$~GeV{#1}}

\begin{document}



\def\ET{\mbox{$E_T$}}
\def\pT{\mbox{$p_T$}}
\def\xT{\mbox{$x_T$}}
\def\kT{\mbox{$k_T$}}
\def\mT{\mbox{$m_T$}}

\def\KThbt{\mbox{$K_T$}}

\def\sqrtsNN{\mbox{$\sqrt{s_\mathrm{_{NN}}}$}}
\def\sqrts{\mbox{$\sqrt{s}$}}
\def\Npart{\mbox{$\mathrm{N}_\mathrm{part}$}}
\def\NpartMean{\mbox{$\langle\Npart\rangle$}}
\def\Nbinary{\mbox{$\mathrm{N}_\mathrm{bin}$}}
\def\NbinaryMean{\mbox{$\langle\Nbinary\rangle$}}

\def\bMean{\mbox{$\langle{b}\rangle$}}

\def\NbinOverNpartMean{\mbox{$\langle\Nbinary/\Npart\rangle$}}

\def\GeVfmCubed{\mbox{$\mathrm{GeV}/\mathrm{fm}^3$}}
\def\GeVperNucleon{\mbox{$\mathrm{GeV}/\mathrm{nucleon}$}}
\def\LuminosityUnits{\mbox{$\mathrm{cm}^{-2}\mathrm{sec}^{-1}$}}

\def\rphihat{\mbox{$r\hat\phi$}}
\def\rhat{\mbox{$\hat{r}$}}
\def\zhat{\mbox{$\hat{z}$}}

\def\pizero{\mbox{$\pi^0$}}
\def\kzeros{\mbox{K$^0_s$}}
\def\kplus{\mbox{K$^+$}}
\def\kminus{\mbox{K$^-$}}

\def\pbar{\mbox{$\bar\mathrm{p}$}}
\def\lam{\mbox{$\Lambda$}}
\def\lambar{\mbox{$\bar\Lambda$}}
\def\xibar{\mbox{$\bar\Xi$}}

\def\SigmaPlus{\mbox{$\Sigma^+$}}
\def\SigmaZero{\mbox{$\Sigma^0$}}

\def\pbarp{\mbox{$\pbar+\mathrm{p}$}}
\def\pp{\mbox{$\mathrm{p}+\mathrm{p}$}}
\def\ePluseMinus{\mbox{$\mathrm{e}^++\mathrm{e}^-$}}

\def\ccbar{\mbox{$\mathrm{c}\bar{\mathrm{c}}$}}
\def\jpsi{\mbox{$\mathrm{J/}\psi$}}
\def\psiPrime{\mbox{$\psi\prime$}}

\def\QBBC{\mbox{$\mathrm{Q}_\mathrm{BBC}$}}
\def\EZDC{\mbox{$\mathrm{E}_\mathrm{ZDC}$}}

\def\Nch{\mbox{$N_{ch}$}}
\def\dNchdeta{\mbox{$\frac{dN_{ch}}{d\eta}$}}
\def\dNchdetaPerNpart{\mbox{$\frac{2}{\langle\mathrm{N}_\mathrm{part}\rangle}\dNchdeta$}}
\def\dNchdetazero{\mbox{$dN_{ch}/d\eta|_{\eta\sim0}$}}

\def\dETdeta{\mbox{$\frac{d\ET}{d\eta}$}}

\def\vtwo{\mbox{$v_2$}}
\def\dedx{\mbox{$dE/dx$}}

\def\GeVc{\mbox{$\mathrm{GeV}/c$}}

\def\sdca{\mbox{$\mathrm{sDCA}$}}
\def\abssdca{\mbox{$|\sdca|$}}

\def\xvert{\mbox{$x_{vert}$}}
\def\yvert{\mbox{$y_{vert}$}}
\def\zvert{\mbox{$z_{vert}$}}
\def\zDCA{\mbox{$z_{DCA}$}}

\def\hplus{\mbox{$\mathrm{h}^+$}}
\def\hminus{\mbox{$\mathrm{h}^-$}}
\def\hphm{\mbox{$(\hplus+\hminus)/2$}}

\def\meanpT{\mbox{$\lt\pT\gt$}}

\def\RAA{\mbox{$R_{AA}(\pT)$}}
\def\RCP{\mbox{$R_{CP}(\pT)$}}
\def\TAA{\mbox{$T_{AA}$}}
\def\RpT{\mbox{$R_{130/200}(\pT)$}}
\def\Reta{\mbox{$R_{\eta}(\pT)$}}

\def\TAB{\mbox{$T_{AB}(\vec{b})$}}

\def\Rsqrts{\mbox{$R_{200/130}(\pT)$}}

\def\cpT{\mbox{$c(\pT)$}}

\def\Qs{\mbox{$Q_s$}}
\def\LamQCD{\mbox{$\Lambda_{QCD}$}}

\def\epsBj{\mbox{$\epsilon_{Bj}$}}

\def\muB{\mbox{$\mu_B$}}

\def\sigmapptot{\mbox{$\sigma^{pp}_{total}$}}
\def\sigmaNNinel{\mbox{$\sigma^{NN}_{inel}$}}
\def\sigmaAAgeom{\mbox{$\sigma^{AuAu}_{geom}$}}
\def\sigmageom{\mbox{$\sigma_{geom}$}}

\def\chisqrndf{\mbox{$\chi^2/ndf$}}

\def\lt{\mbox{$<$}}
\def\gt{\mbox{$>$}}

\def\invarXsection{\mbox{$\frac{1}{2\pi{p}_T}\frac{dN}{dp_T}$}}

\def\IntLumi{\mbox{$\int\mathcal{L}dt$}}

\def \mpt   {$\langle p_T \rangle$}
\def \pt   {$p_T$}
\def \pp	{p+p }
\def \dau	{d+Au }
\def \pau	{p+Au }
\def \auau	{Au+Au }
\def \cucu	{Cu+Cu }
\def \sial	{Si+Al }
\def \pbpb	{Pb+Pb }
\def \raa	{$R_{AB}$ }
\def \v2        {$v_2$ }
\newcommand{ \rts }{$\sqrt{s_{_{\rm NN}}}$ }
\def \npart {$\langle N_{part} \rangle$ }

\newcommand{\deta}{$\Delta\eta$ }
\newcommand{\dphi}{$\Delta\phi$ }
\newcommand{\dphiabs}{$|\Delta\phi|$ }
\newcommand{\detano}{$\Delta\eta$}
\newcommand{\dphino}{$\Delta\phi$}
\newcommand{\detaj}{$\Delta\eta(J)$ }
\newcommand{\dphij}{$\Delta\phi(J)$ }
\newcommand{\dphijr}{$\Delta\phi(J+R)$ }
\newcommand{\detajno}{$\Delta\eta(J)$}
\newcommand{\dphijno}{$\Delta\phi(J)$}
\newcommand{\dphijrno}{$\Delta\phi(J+R)$}
\newcommand{\pttrig}{$p_{t}^{trig}$ }
\newcommand{\ptass}{$p_{t}^{assoc}$ }
\newcommand{\pttrigno}{$p_{t}^{trig}$}
\newcommand{\ptassno}{$p_{t}^{assoc}$}
\newcommand{\ks}{$\mathrm{K}^{0}_{S}$ }


\newcommand{\ttbs}{\char'134}
\newcommand{\AmS}{{\protect\the\textfont2
  A\kern-.1667em\lower.5ex\hbox{M}\kern-.125emS}}
\hyphenation{author another created financial paper re-commend-ed}
\newcommand{ \mt }{$m_t$ }
\def \mpt   {$\langle p_T \rangle$}
\def \hijb  {HIJING/B$\overline{\rm B}$}
\def \hij   {HIJING}

\def \npart {$\langle N_{part} \rangle$ }
\def \mbeta {$\langle \beta_t \rangle$}
\def \ppbar {p+$\overline{\rm p}$}
\newcommand{ \pbp }{ \overline{p}/p }

\hyphenation{nuc-leon spec-tro-me-ter qua-dru-pole inter-act-ions}

\begin{center}
{\Huge \bf Hadron Production in Heavy Ion Collisions }\\
\vspace{20 mm}

{\huge \bf Helmut Oeschler} \\
{\huge \bf Institut f\"ur Kernphysik} \\
{\huge \bf Darmstadt University of Technology} \\
{\huge \bf 64289 Darmstadt, Germany} \\
\vspace{10 mm}
{\huge \bf Hans Georg Ritter and Nu Xu} \\
{\huge \bf Nuclear Science Division} \\
{\huge \bf Lawrence Berkeley National Laboratory} \\
{\huge \bf  Berkeley, Ca 94720, USA} \\
\vspace{5 mm}

\end{center}


\tableofcontents

\newpage

\section{Introduction}
\label{introduction}

Heavy ion collisions are an ideal tool to explore the QCD phase
diagram.  The goal is to study the equation of state (EOS) and to
search for possible in-medium modifications of hadrons. By varying
the collision energy a variety of regimes with their specific
physics  interest can be studied. At energies of a few GeV per
nucleon, the regime where experiments were performed first at the
Berkeley Bevalac and later at the Schwer-Ionen-Synchrotron (SIS)
at GSI in Darmstadt, we study the equation of state of dense
nuclear matter and try to identify in-medium modifications of
hadrons. Towards higher energies, the regime of the Alternating
Gradient Synchrotron (AGS) at the Brookhaven National Laboratory
(BNL), the Super-Proton Synchrotron (SPS) at CERN, and the
Relativistic Heavy Ion Collider (RHIC) at BNL, we expect to
produce a new state of matter, the Quark-Gluon Plasma (QGP). The
physics goal is to identify the QGP and to study its properties.

By varying the energy, different forms of matter are produced. At
low energies we study dense nuclear matter, similar to the type of
matter neutron stars are made of.  As the energy is increased the
main constituents of the matter will change. Baryon excitations
will become more prevalent (resonance matter). Eventually we
produce deconfined partonic matter that is thought to be in the
core of neutron stars and that existed in the early universe.

At low energies a great variety of collective effects is observed
and a rather good understanding of the particle production has
been achieved, especially that of the most abundantly produced
pions and kaons. Many observations can be interpreted as
time-ordered emission of various particle species. It is possible
to determine, albeit model dependent, the equation of state of
nuclear matter. We also have seen indications, that the kaon mass,
especially the mass of the K$^+$, might be modified by the medium
created in heavy ion collisions.

At AGS energies and above, emphasis shifts towards different
aspects. Lattice QCD calculations~\cite{satzqm02,karsch} predict
the transition between a Quark-Gluon Plasma and a hadronic state
at a critical temperature, $T_c$, of about 150 to 190 MeV at
vanishing baryon density. The energy density at the transition
point is about $1.0$ GeV/fm$^3$. It is generally assumed that
chiral symmetry restoration happens simultaneously
\cite{satzqm02}.

In the high-energy regime, especially at RHIC, a rich field of
phenomena~\cite{STAR_white} has revealed itself.
Hot and dense matter with very strong collectivity has been
created. There are indications that collectivity develops at
the parton level, i.e.~at a very early stage of the collision,
when the constituents are partons rather than hadrons. Signs of
pressure driven collective effects are  our main tool for the
study of the EOS. There are also strong indications that in the
presence of a medium hadronization occurs through the process of
quark coalescence and not through quark fragmentation,
the process dominant for high-energy p+p reactions.

We limit this report to the study of hadrons emitted in heavy ion
reactions. The report is divided into two parts. The
first part describes the phenomena observed from hadrons produced
at low energies, whereas the second part concentrates on the
search for signs of a partonic state at high energies.

\section{Hadron Production below 2 A GeV}
\label{low_energy}

In the discussion of particle production in heavy ion collisions
below 2 \AGeV we focus  on the most abundantly produced particles,
pions and kaons. Pion production is the dominant channel for
particle emission, while kaon emission is a rare process. Kaons
are produced during the high-density phase. Therefore, kaons are
considered as ideal probes for the hot and dense fireball.

In fixed-target p+p collisions the threshold for pion production
is 0.29 GeV, for K$^+$ it is 1.58 GeV and for K$^-$ it is 2.5 GeV.
The threshold for K$^+$ is lower than that for K$^-$, because the
energy balance has only to account for the excitation of a nucleon
to a  $\Lambda$ in the case of  K$^+$, whereas together with the \km a
\kp has to be produced to conserve strangeness.

\begin{figure}[h]
\centerline{\includegraphics[width=.5\textwidth]{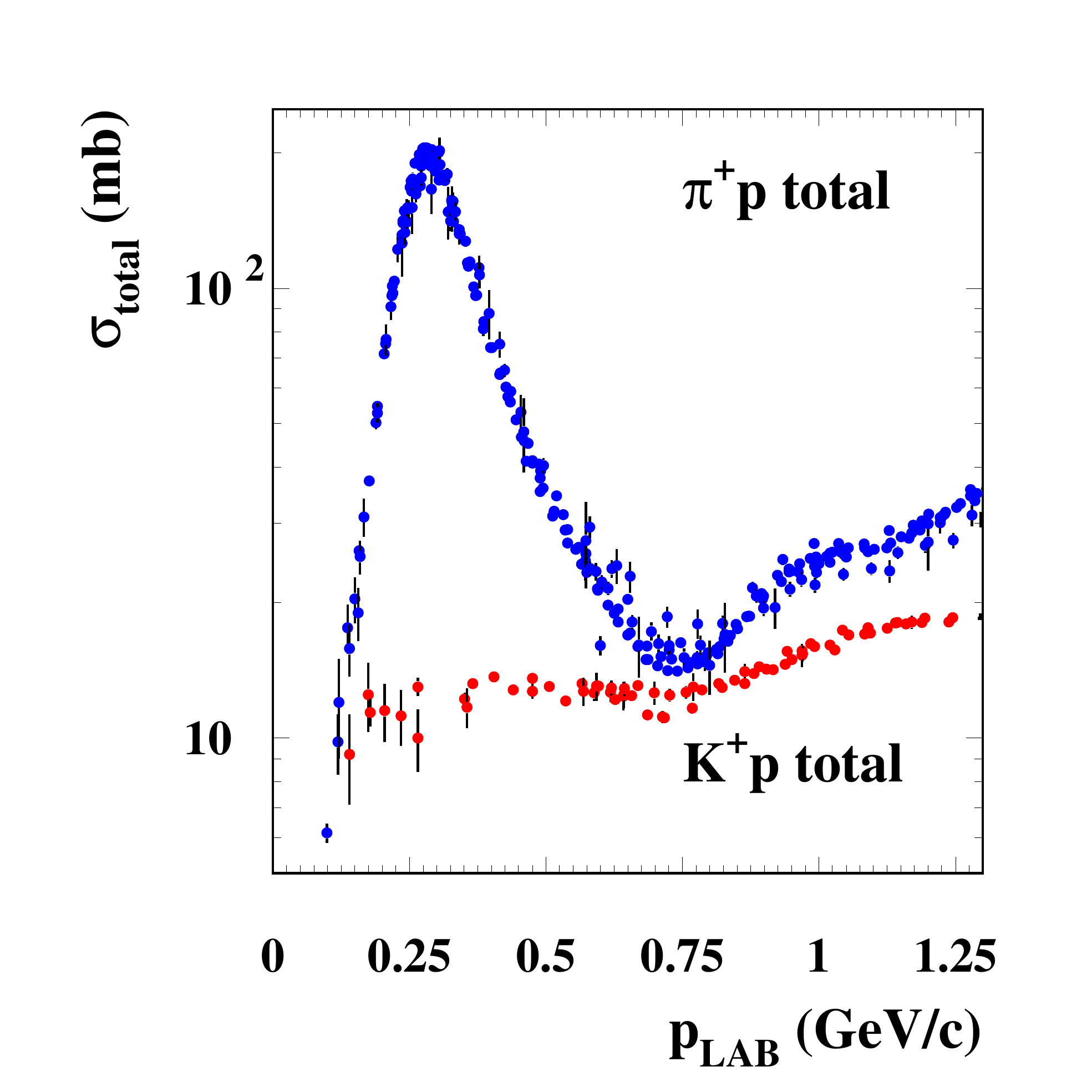}
\vspace*{-.4cm} }\caption{Elementary cross sections for K$^+{\rm
p}$ and $\pi^+{\rm p}$ interactions~\cite{PDG}. } \label{pdg}
\end{figure}

The reaction dynamics  of pions and kaons is quite different. The
pion-nucleon cross section, shown in Figure~\ref{pdg}, is large
and thus pions are continuously absorbed through the
$\Delta$-resonance and re-emitted by its decay. This creation and
disappearance can occur during the entire time evolution of the
collision. The K$^+$-nucleon interaction cross section, also shown
in Figure~\ref{pdg}, on the other hand, is small due to
strangeness and energy conservation. There are no partners to
react with and only elastic scattering and charge exchange can
occur. Therefore, \kp mesons  are expected to leave the
interaction zone early. These facts have led to the suggestion of
time-ordered emission of these two species~\cite{Nagamiya_82a}.
The creation of \km requires even higher energies. The dominant
production mechanism is strangeness exchange. The interaction
probability of \km with nuclear matter is much higher than that of
K$^+$ and the mean free path much lower. As a consequence, \km
cannot leave the reaction zone undisturbed and they are emitted
later.

\begin{figure}[]
\centerline{\includegraphics[width=.5\textwidth]{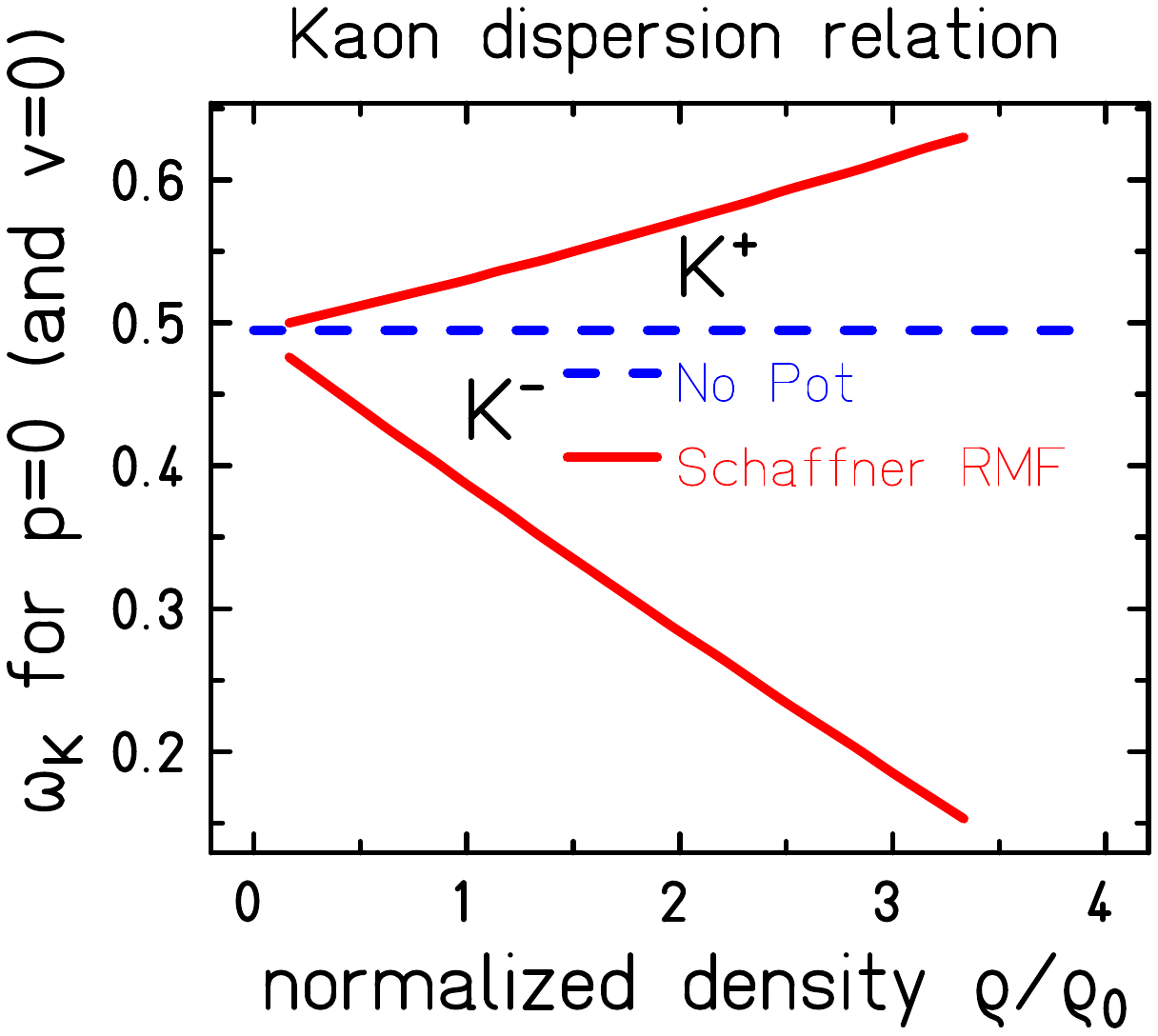}}
\caption{The effective mass of \kp  and of  \km resulting from the
KN potential calculated within a relativistic mean field
theory~\cite{Schaffner} as a function of the nuclear
density~\cite{iqmd_PR}.} \label{knpot}
\end{figure}

The properties of kaons are expected to be changed inside the
nuclear medium~\cite{Kaplan, Schaffner}. The scalar part of the KN
potential of both kaon species is slightly attractive while the
vector part acts in opposite direction. Thus, the \kp N potential
is slightly repulsive and the \km N potential is strongly
attractive. Potentials increase with density as shown in
Figure~\ref{knpot}. Kaons emitted in heavy ion collisions are an
ideal tool to study this hypothesis.

Kaons are also used to extract key parameters of the nuclear
equation of state, like the nuclear compressibility. A large
fraction of the work to extract the parameters of the EOS has been
carried out through flow studies of the nucleons. A separate
report in this Volume~\cite{Aich_Rev} describes these efforts in
detail.

Pion distributions were measured at the Bevalac in light symmetric
systems ($A<40$) and in asymmetric systems
\cite{Nagamiya_81,Nagamiya_82,Wolf_82,Brockmann_84} and later also
in the La+La system~\cite{Hayashi_88}.
In these early measurements the slope of the high-$p_T$ part of
the spectra was taken as a measure for the freeze-out temperature,
$T_{fo}$. This is a good approximation since in the blast-wave
formalism~\cite{Siemens_79} the radial expansion does not strongly
modify the spectrum due to the low mass of the pions. The pion
multiplicity as a function of beam energy was measured by the
Streamer Chamber Collaboration~\cite{Harris_87}. The difference
between the measured pion yield and the yield predicted by cascade
calculations~\cite{Cugnon_81} has been taken as an early attempt
to determine the equation of state of nuclear
matter~\cite{Harris_85}.

Positive kaons were measured early in the Bevalac
program~\cite{Schnetzer_82,Schnetzer_89,Shor_89}. However, the
statistics and systematics were not good enough to reconstruct a
complete emission pattern the way it was done for pions. At Bevalac
energies the production of negatively charged  kaons is
sub-threshold. The early measurements lack good
statistics~\cite{Shor_89,Shor_82}, which adds to the problem to
explain the observed phenomena close to the threshold where the
production cross section varies rapidly with energy. Lambda and
antiproton production~\cite{Shor_82,Harris_81,Carroll_89} suffers
from similar problems.
A detailed overview of  hadron production in
high-energy nucleus-nucleus collisions at the Bevalac can be found
in a review paper~\cite{Stock_86}.

Systematic, high-statistics measurements of pion and kaon
production became feasible with the advent of the SIS accelerator
at GSI. Two experiments, KaoS~\cite{Senger_93}, a dedicated
experiment for the measurement of $K$ mesons, and
FOPI~\cite{Gobbi_93}, a large acceptance, multi-purpose detector,
produced a solid body of beautiful results.
References~\cite{reisdorf-pionen} and \cite{KaoS} give a detailed
summary of pion and kaon production at SIS.

Cascade codes are an ideal tool for the study of the mechanism of
hadron production. At the Bevalac, pion
production~\cite{Cugnon_81} and kaon production~\cite{Randrup_80}
was implemented in some models.  Theory evolution not being part
of this review, we limit ourselves to studying the production
mechanism with the help (by means) of one particular model, the
Isospin Quantum Molecular Dynamics (IQMD)
model~\cite{iqmd_PR,iqmd}. Reference~\cite{Fuchs_Rev_06} reviews
the theory development mainly concentrating on the nuclear
equation of state. We focus on reviewing multiplicities and
spectral distributions.

\subsection{Multiplicities}

 \label{mult}

Pions are the most abundantly produced particles. They are created
in individual nucleon-nucleon collisions via the
$\Delta$-resonance through the reaction $NN \rightarrow N\Delta
\rightarrow NN\pi$. Pions interact strongly with nuclear matter by
forming baryonic resonances, e.g.~ $\pi + $N$ \rightarrow
\Delta$. The resonances themselves decay mainly by pion emission.
Therefore, pions are expected to leave the collision zone over a
long time span, mainly at a late stage of the collision when the
system has expanded and cooled. For an experimental study of
$\Delta$-production we refer to Reference~\cite{hong_fopi}.

\begin{figure}[h]
\centerline{\epsfig{file=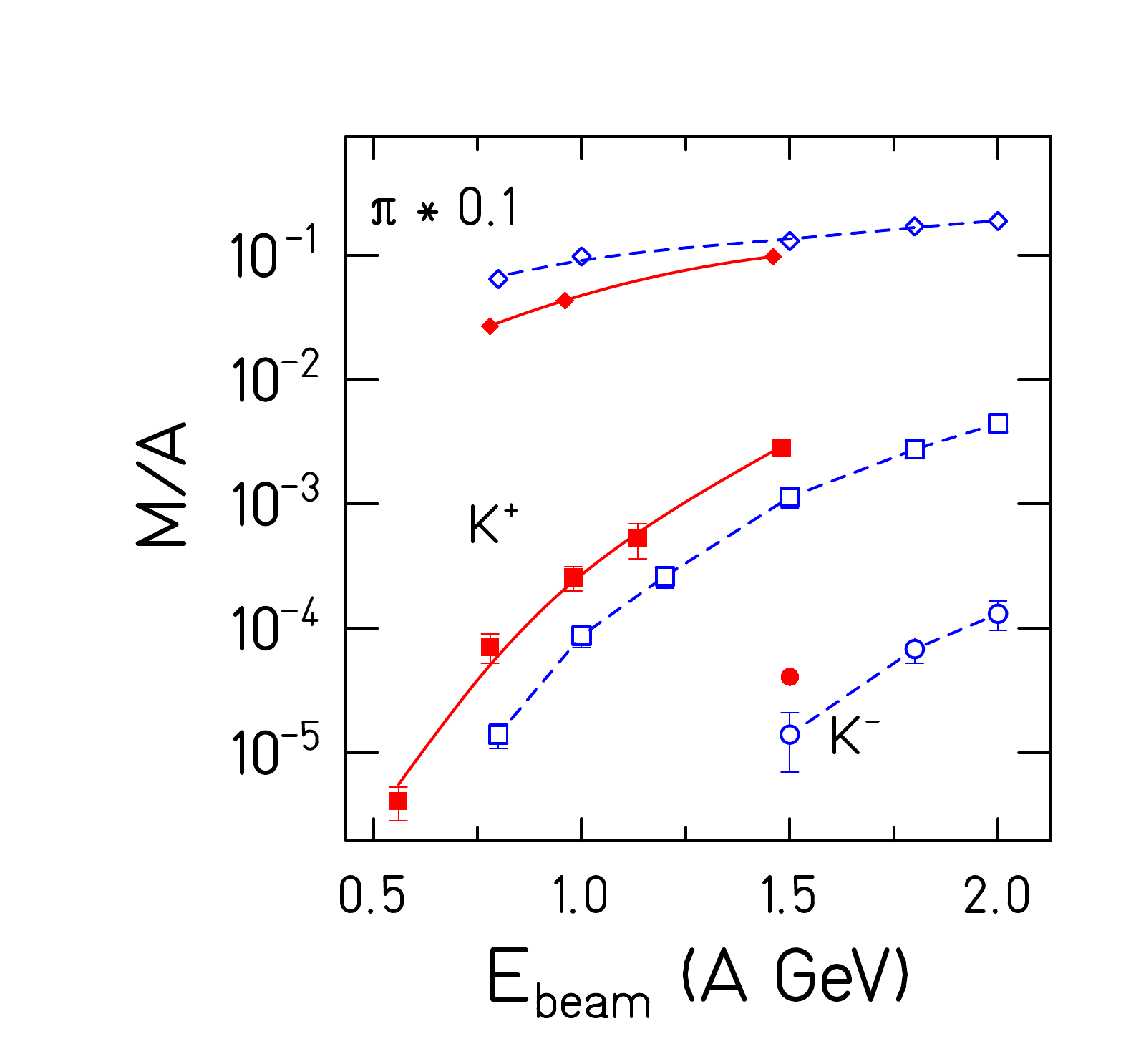,width=9cm}}
 \caption{ Multiplicity per mass number $A$ (of one collision partner)
 for pions (diamonds, sum of all pion species), of K$^+$ (squares), and of K$^-$ (circles)
  for Au+Au (filled symbols) and for C+C (open symbols) as a
    function of the beam energy.  Values are taken from   \cite{KaoS,sturm}. }
\label{k_sigma_Energy}
\end{figure}

 The pion multiplicity increases slowly as a
function of the collision energy as shown in
Figure~\ref{k_sigma_Energy} for inclusive collisions of a light
(C+C) and a heavy (Au+Au) system. The ordinate $M/A$, the
multiplicity per mass number $A$ of one collision partner,
accounts for the system size. The yield is higher for the lighter
system. This reflects absorption of pions and holds only at these low
energies. The kinematical conditions for \pip and \pim
are the same, the only difference is given by the $N/Z$ ratio of
the interacting nuclei. This effect is avoided here by plotting
the sum of all three pion species.

Figure~\ref{k_sigma_Energy} also shows the K$^+$ and
K$^-$ production yields for the two systems. As expected, the K$^+$ multiplicity is higher than the K$^-$ multiplicity.
 The pion multiplicity is higher in the lighter system
while for K$^+$ production the inverse observation is made.

In central \mbox{Au+Au} collisions, densities of two to three
times normal nuclear matter density may be
reached~\cite{iqmd_PR,iqmd,Fuchs_Rev_06}. A sensitive probe to
test such conditions is the production of strange hadrons at or
below the production threshold.

\begin{figure}[]
\centerline{\epsfig{file=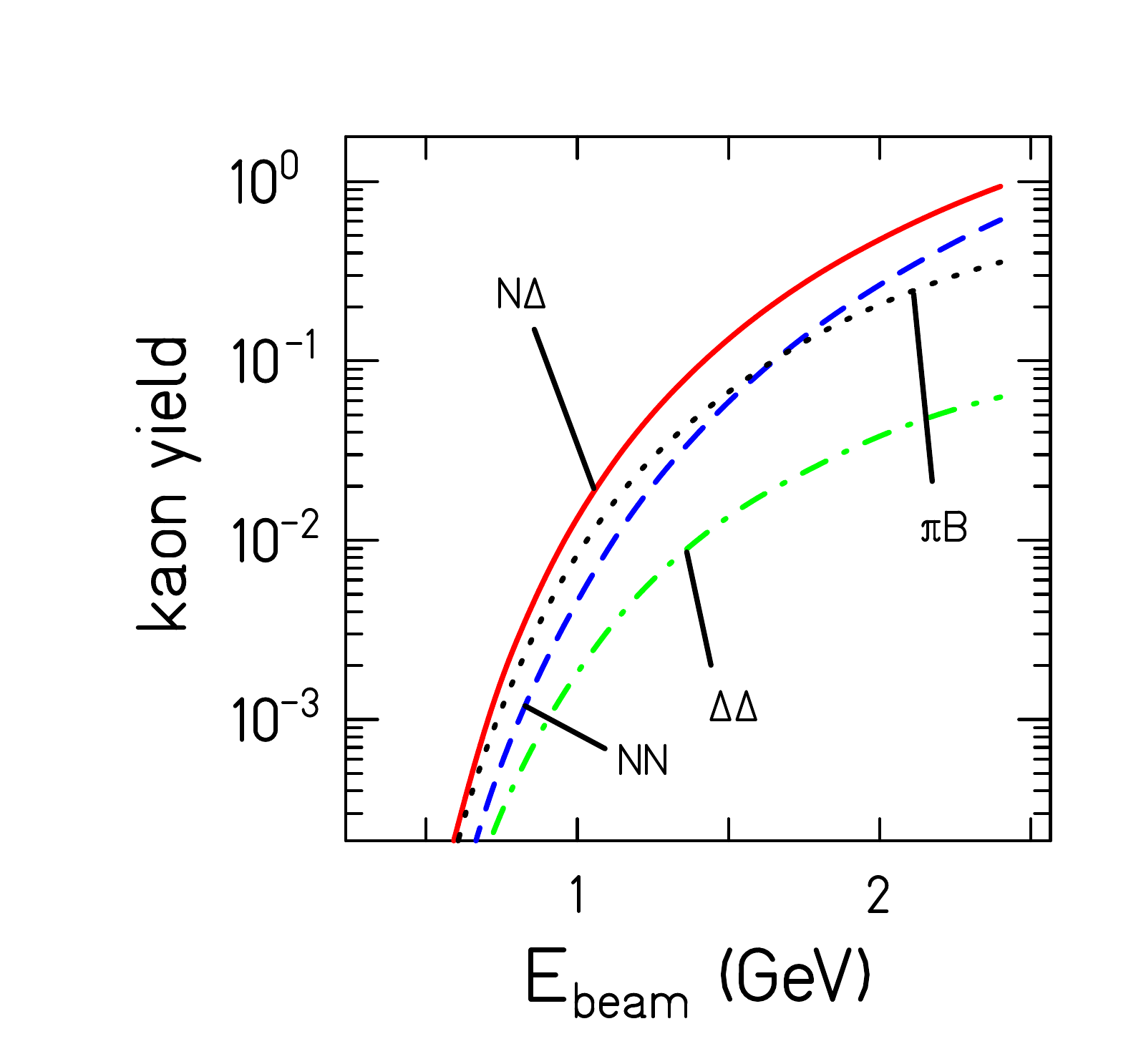,width=0.5\textwidth}}
\caption{Contribution of different production channels to the
K$^+$ yield in central Au+Au collisions as a function of the beam
energy~\cite{iqmd_PR}.} \label{AuAu_excitation_function}
\end{figure}

The key mechanism for K$^+$ production
close to the threshold is a multi-step process where the energy
necessary for production is accumulated in intermediate
resonances.  Higher density increases the number of these
collisions. Especially second generation collisions with
sufficiently high relative momentum to create a K$^+$ occur most
frequently during the high-density phase.
Figure~\ref{AuAu_excitation_function} shows the contribution
of different production channels to the kaon yield in central
Au+Au collisions according to IQMD calculations~\cite{iqmd_PR}.
$\Delta \rm{N} $  collisions dominate even above the corresponding
NN threshold.

The K$^-$ production process is quite different from K$^+$ and the
production threshold is much higher. The strangeness-exchange
reaction $ \pi \rm{Y} \rightleftharpoons \rm{K}^- \rm{N}$, with Y
being $\Lambda$ or $\Sigma$, represents an additional production
possibility, as suggested by Ko~\cite{Ko_84} and demonstrated in
References~\cite{Foerster_03,Oeschler_00,Hartnack_03}. It has a
large cross section. The inverse channel causes the produced K$^-$
to be absorbed~\cite{Hartnack_03}. As in the pion case, the
succession of absorption and creation causes the K$^-$ emission to
be mainly in the late stage of the reaction.

\begin{figure}
\epsfig{file=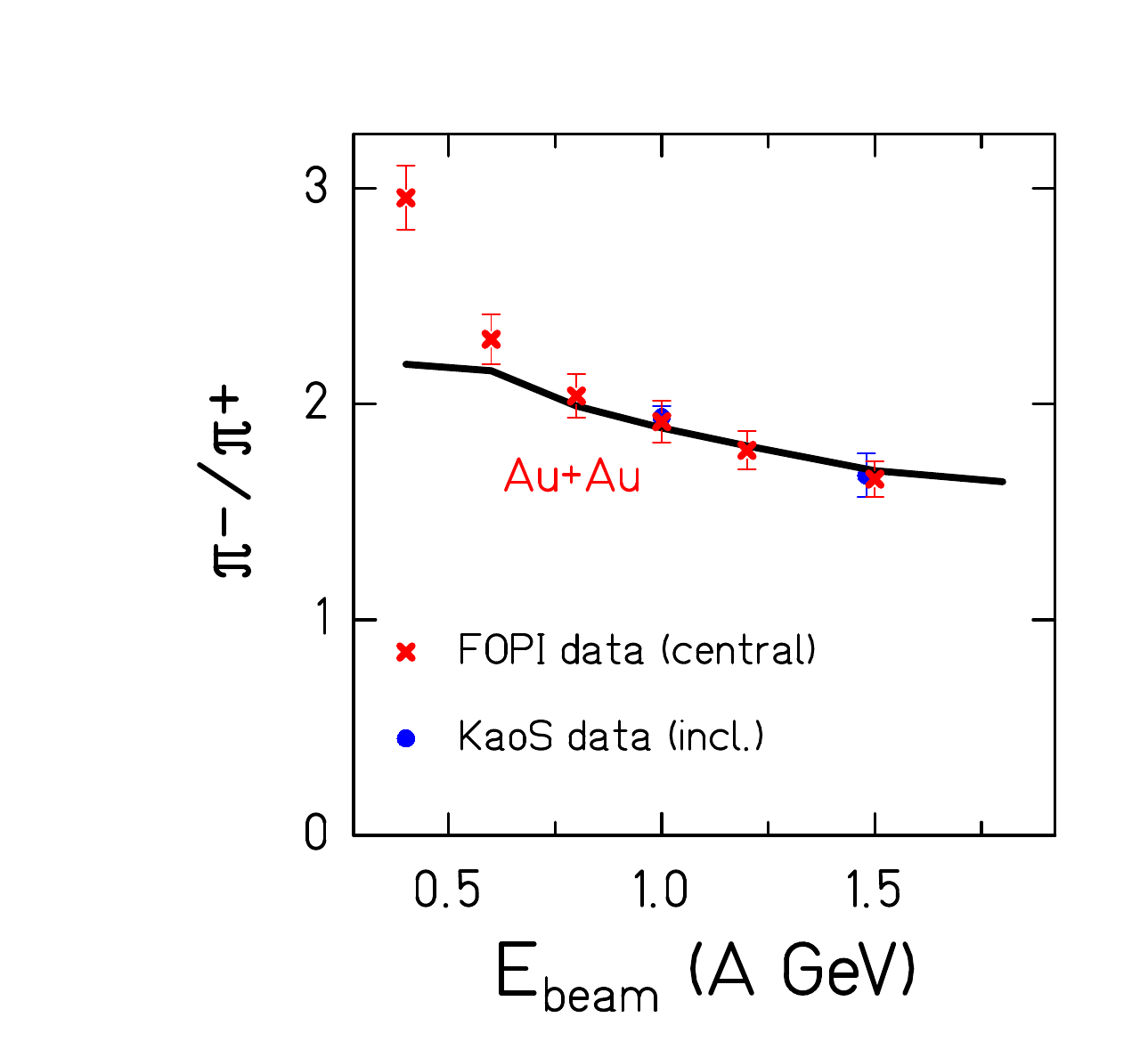,width=0.49\textwidth}
\epsfig{file=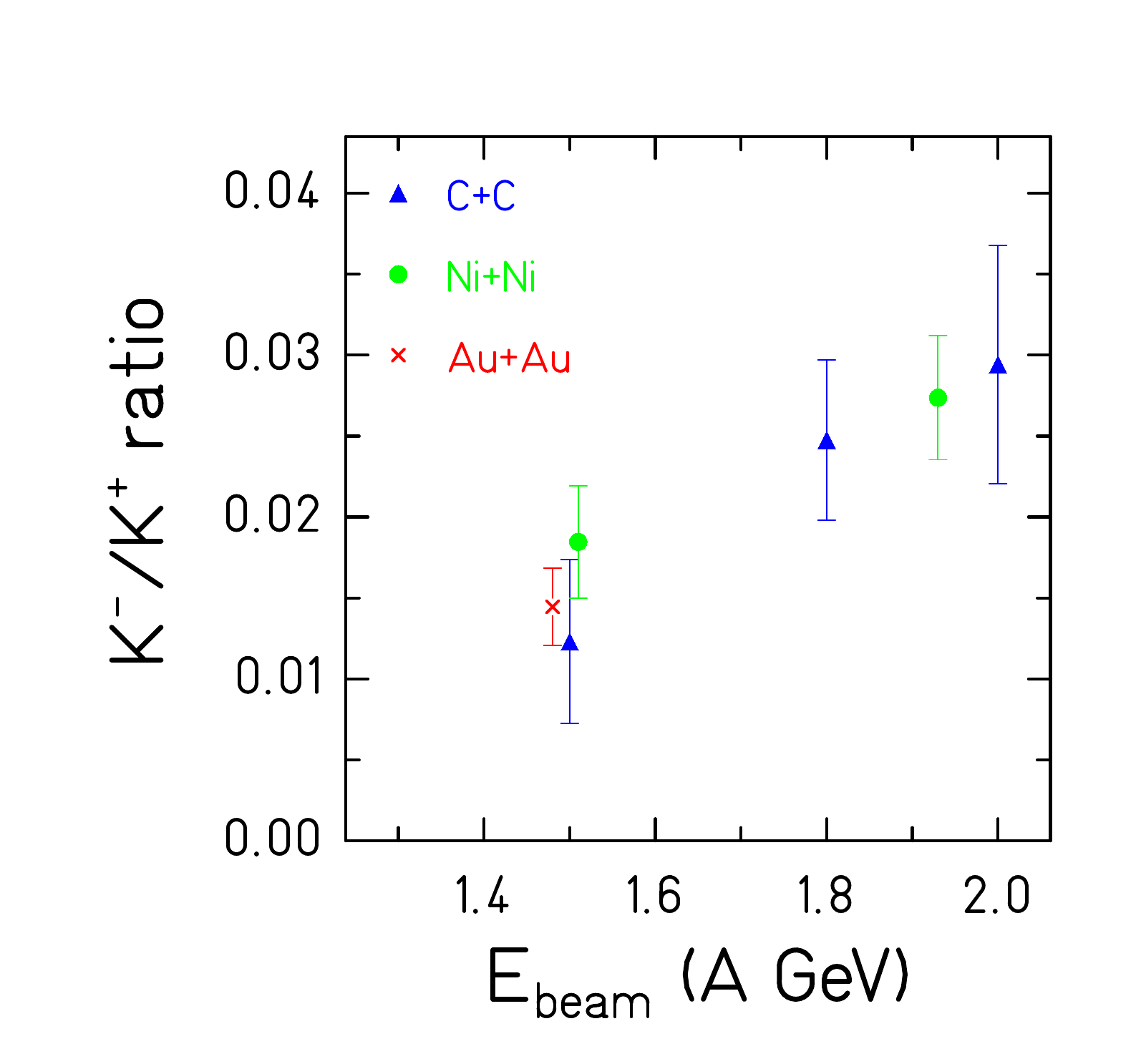,width=0.49\textwidth}
\caption{The $\pi^-/\pi^+$ ratio (left panel) and the  K$^-$/K$^+$
ratio (right panel) as a function of the beam energy. The data are
taken from References~\cite{reisdorf-pionen} and \cite{KaoS}. The
solid line refers to IQMD calculations~\cite{iqmd_PR}.}
\label{pim-pip-ratio}
\end{figure}

The $\pi^-/\pi^+$ ratio reflects the $N/Z$ ratio of the colliding
system. Assuming $\pi$ production via the $\Delta$-resonance,
the ratio is 1.95 (isobar value)
for Au+Au collisions. The left panel of
Figure~\ref{pim-pip-ratio} shows the $\pi^-/\pi^+$ ratio as  a
function of beam energy. The ratio increases with decreasing
energy, exceeding even the isobar value. IQMD calculations~\cite{iqmd_PR} reproduce the trend but fail to describe the
strong rise at very low incident energies. At very high
energies the value reaches unity due to charge
conservation~\cite{STARratio}.

The K$^-$/K$^+$ ratio, shown in the right panel of
Figure~\ref{pim-pip-ratio}, rises strongly with increasing energy.
This reflects the different production thresholds and kinematics
for the two particles.  The three systems studied here show quite
similar values. The fact that different densities are reached in
the different systems apparently does not matter. This trend is
well described by transport models where strangeness exchange is
the dominant channel and by statistical
models~\cite{Cleymans_99}. Again, at very high energies the value
reaches unity~\cite{STARpaper}.

In the following we study the multiplicities of pions and kaons as
a function of the mass of the collision system, $A+A$, for inclusive
reactions and as a function of the number of participants, $A_{part}$, for impact parameter selected reactions.

\begin{figure}
\centerline{\epsfig{file=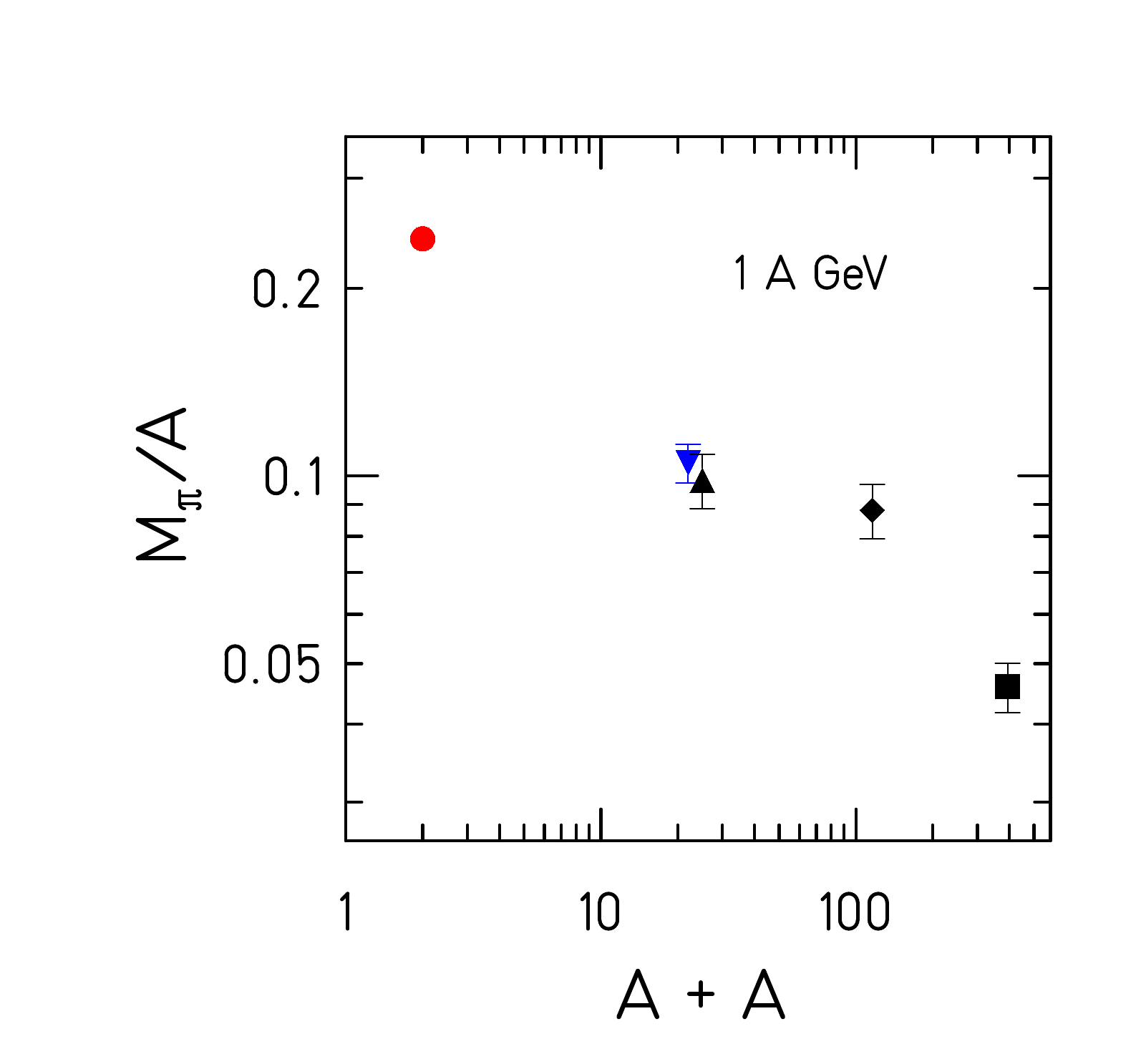,width=0.49\textwidth}}
\caption{Pion (sum of $\pi^+, \pi^0, \pi^-$)  multiplicities per
$A$ for inclusive collisions of p+p (divided by 2), C+C, Ni+Ni and
Au+Au reactions as a function of $A+A$, the total mass of the
system. The data are taken from
References~\cite{reisdorf-pionen,Averbeck_03,Sturm_diss}.}
\label{kaon_pion_apart}
\end{figure}

\begin{figure}
\centerline{\epsfig{file=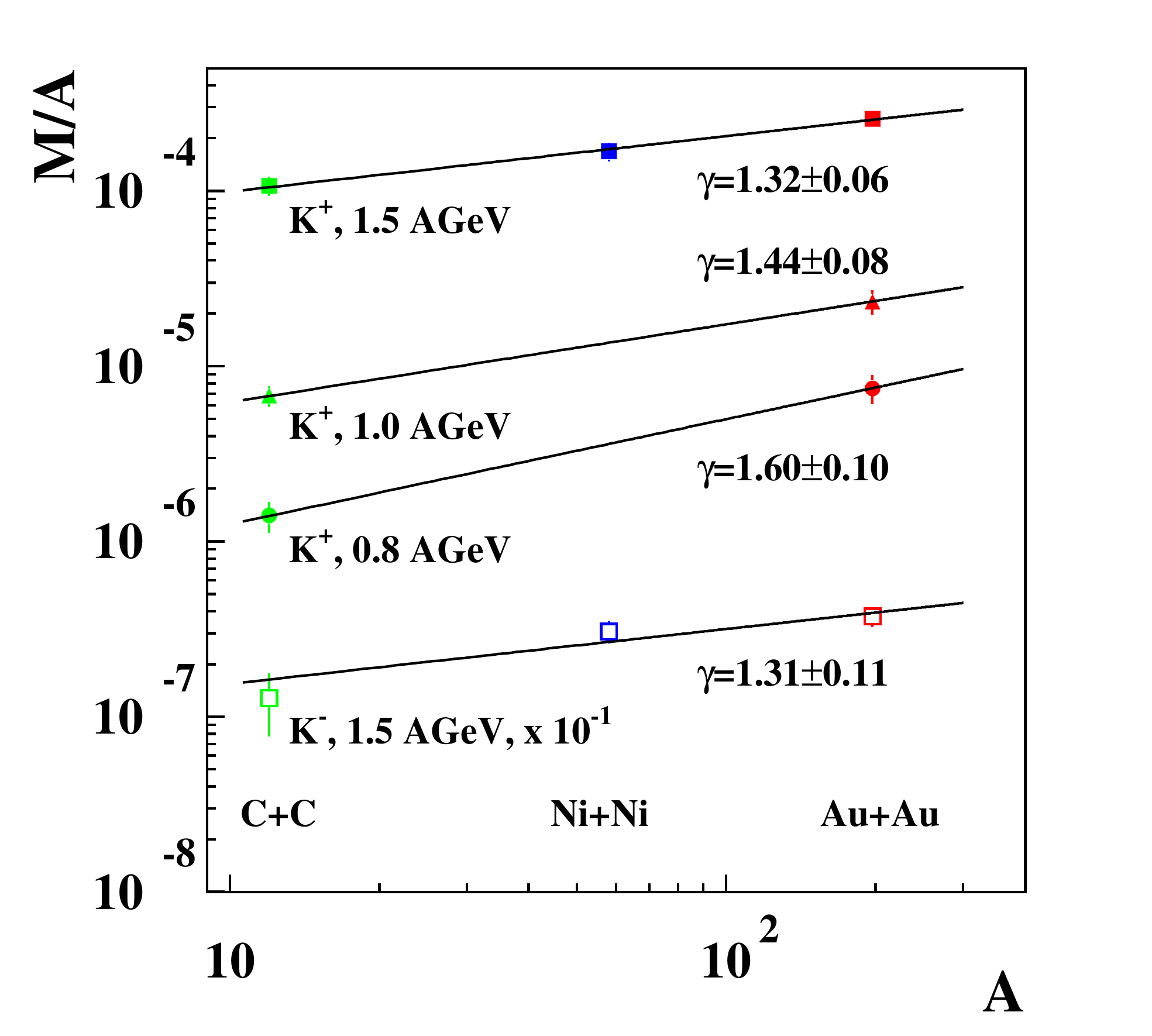,width=8cm}} \caption{
Multiplicities per mass number $M/A$ as a function of $A$ for
K$^+$ (full symbols) and for K$^-$ (open symbols) for inclusive
\mbox{C+C}, \mbox{Ni+Ni}, and \mbox{Au+Au} reactions. The lines
represent the function  $M \sim A^{\gamma}$ fitted to the data.
From \cite{KaoS}.} \label{k_sigma_a}
\end{figure}

Figure~\ref{kaon_pion_apart} shows the inclusive multiplicity per
mass number, $M/A$, for pions as a function of $A+A$ at 1 \AGeV
incident energy. The pion multiplicity per $A$ decreases strongly
with $A$. This is in strong contrast to the behavior of \kp
production, shown in Figure~\ref{k_sigma_a}. At 1 $A$ GeV, the
normalized kaon yield rises by a factor of about 3.
Figure~\ref{k_sigma_a} also shows the variation of multiplicities
of K$^+$ mesons  per $A$ with incident energy as well as those of
K$^-$ mesons at 1.5~$A$~GeV. The lines are functions $M \sim
A^{\gamma}$ fitted to the data with the resulting values for
$\gamma$ given in the figure. For K$^+$ production the extracted
values of $\gamma$ decrease with incident energy. This reflects
the decreasing influence of intermediate energy storage via
$\Delta$ as less multiple collisions are needed. Considering the
much higher threshold for K$^-$ production, one would expect at
the same incident energy a much stronger rise. However, at 1.5
\AGeV the values of $\gamma$ for K$^+$ and for K$^-$ production
are about equal, demonstrating that K$^-$ production and K$^+$
production are strongly correlated.

\begin{figure}[]
\centerline{\epsfig{file=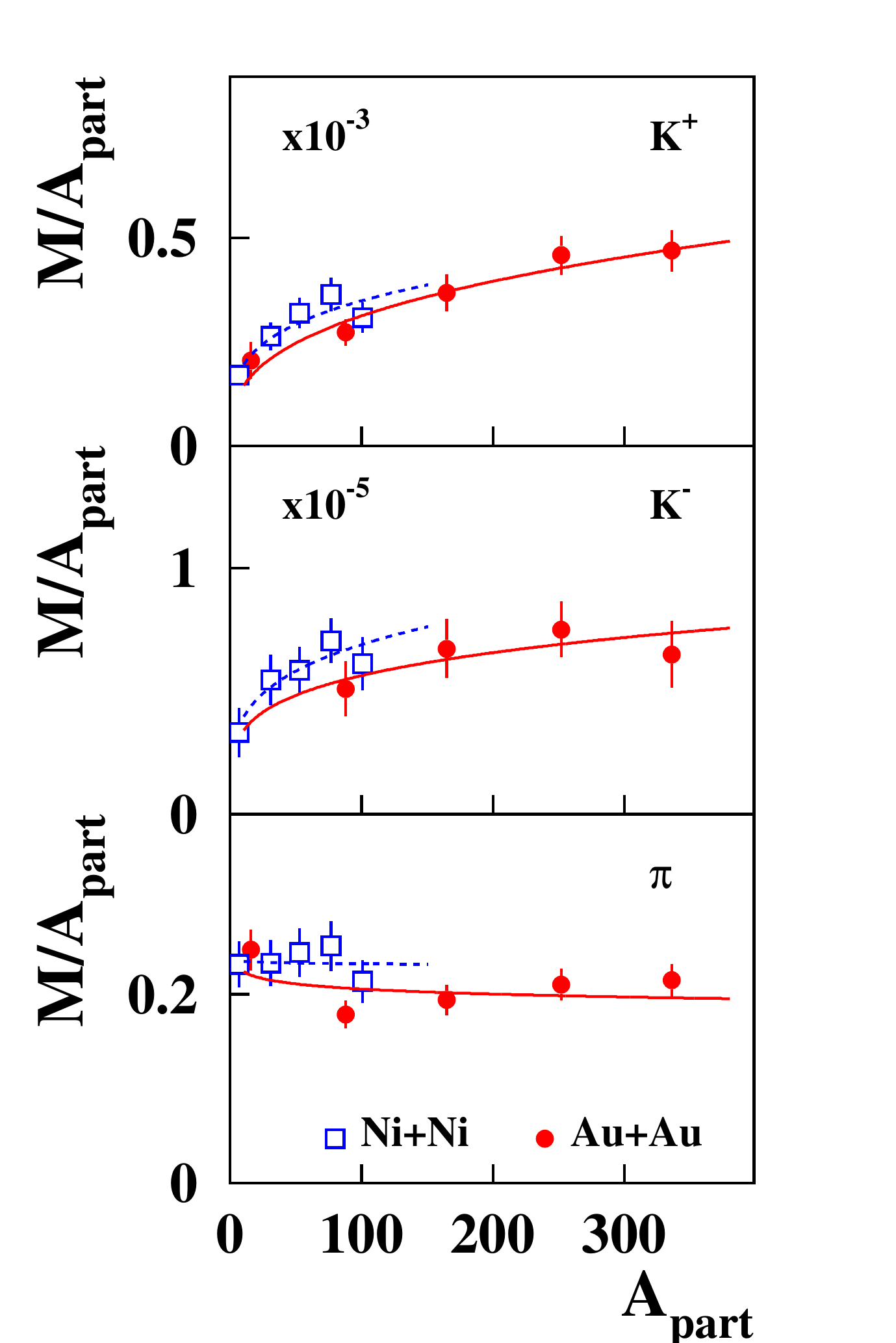,width=7cm}}
 \caption{
Dependence of the multiplicities of K$^+$ (upper panel) and of
K$^-$ mesons (middle panel) as well as of pions (lower panel) on
$A_{part}$. Full symbols denote \mbox{Au+Au}, open symbols
\mbox{Ni+Ni}, both at 1.5 \AGeV. The lines are functions $M \sim
A^{\alpha}_{part}$ fitted to the data separately for
\mbox{Au+Au} (solid lines) and \mbox{Ni+Ni} (dashed lines).
  The data have been measured at $\theta_{\rm{lab}} = 40^{\circ}$.
  Taken from~\cite{KaoS}.}
\label{Au_ratio}
\end{figure}

Figure~\ref{Au_ratio} shows $M/A_{part}$ for Ni+Ni and Au+Au
collisions at 1.5 \AGeV as a function of $A_{part}$. It
demonstrates that the multiplicities of both kaon species exhibit
the same rise with the number of participating nucleons  despite the fact that the thresholds for the production of
the two particle species are very different.
This observation again confirms that K$^+$ and K$^-$ productions
are correlated.
As is also shown in Figure~\ref{Au_ratio}, the pion multiplicity
per \Apart as a function of \Apart is rather flat. This
observation has been made at several incident
energies~\cite{Harris_87, Schwalb_94, Muentz_95}.

It is important to distinguish the trends observed in the quantity
$M/A$ for inclusive reactions and $M/A_{part}$ studied for
different centralities. While for kaons these trends are quite
similar as demonstrated in Figures~\ref{k_sigma_a} and
\ref{Au_ratio}, it is not the case for pions. While $M/A_{part}$
is flat, $M/A$ decreases when comparing different systems (see
Figure~\ref{kaon_pion_apart}). Consequently, $M/A_{part}$ for
different systems gives different values. Results from inclusive
and from centrality-selected studies can not easily be compared.

The similar rise of both, K$^+$ and K$^-$, as a function of the
collision centrality shown in Figure~\ref{Au_ratio} and as a
function of the system size in Figure~\ref{k_sigma_a}, suggests
that the production mechanisms of the two kaon species is
correlated. The K$^-$ in heavy ion collisions at SIS energies are
mainly produced via strangeness-exchange. On the other hand,
strangeness has to be conserved when producing these hyperons and
the energetically most favorable way is to produce them together
with K$^+$ (or K$^0$) mesons. Thus the production of K$^+$ and of
K$^-$ mesons is coupled via the strangeness-exchange reaction and
the K$^-$ inherits the same dependence on the system size and on
the collision centrality. In Reference~\cite{Cleymans_04} it is
argued that the strangeness-exchange channel reaches chemical
equilibrium. Thus
 the \mbox{K$^-$/K$^+$} ratio is proportional to
the pion density,
both at SIS and AGS energies.

As a consequence of Figure~\ref{Au_ratio}, the ratio
\mbox{K$^-$/K$^+$} as a function of $A_{part}$ is constant and it
is the same for \mbox{Au+Au} and for \mbox{Ni+Ni}. Transport
models having strangeness exchange as dominant production channel
for K$^-$, describe this trend fairly well~\cite{Hartnack_03}. The
statistical model reproduces the measured values using nominal
masses for all particles~\cite{Cleymans_99}. This model predicts a
value independent of centrality and of system size. At these low
incident energies, a canonical treatment is required leading to
multiplicities of strange particles depending on the size of the
fireball (correlation volume). In this approach the size
dependence of both, \kp and K$^-$, is the same and a constant
\mbox{K$^-$/K$^+$} ratio results.

\subsection{Spectra}

The spectral distributions of the different particles in general
contain information about the freeze-out condition and about the
collective expansion of the system created in heavy ion
collisions.

\begin{figure}
\centerline{\epsfig{file=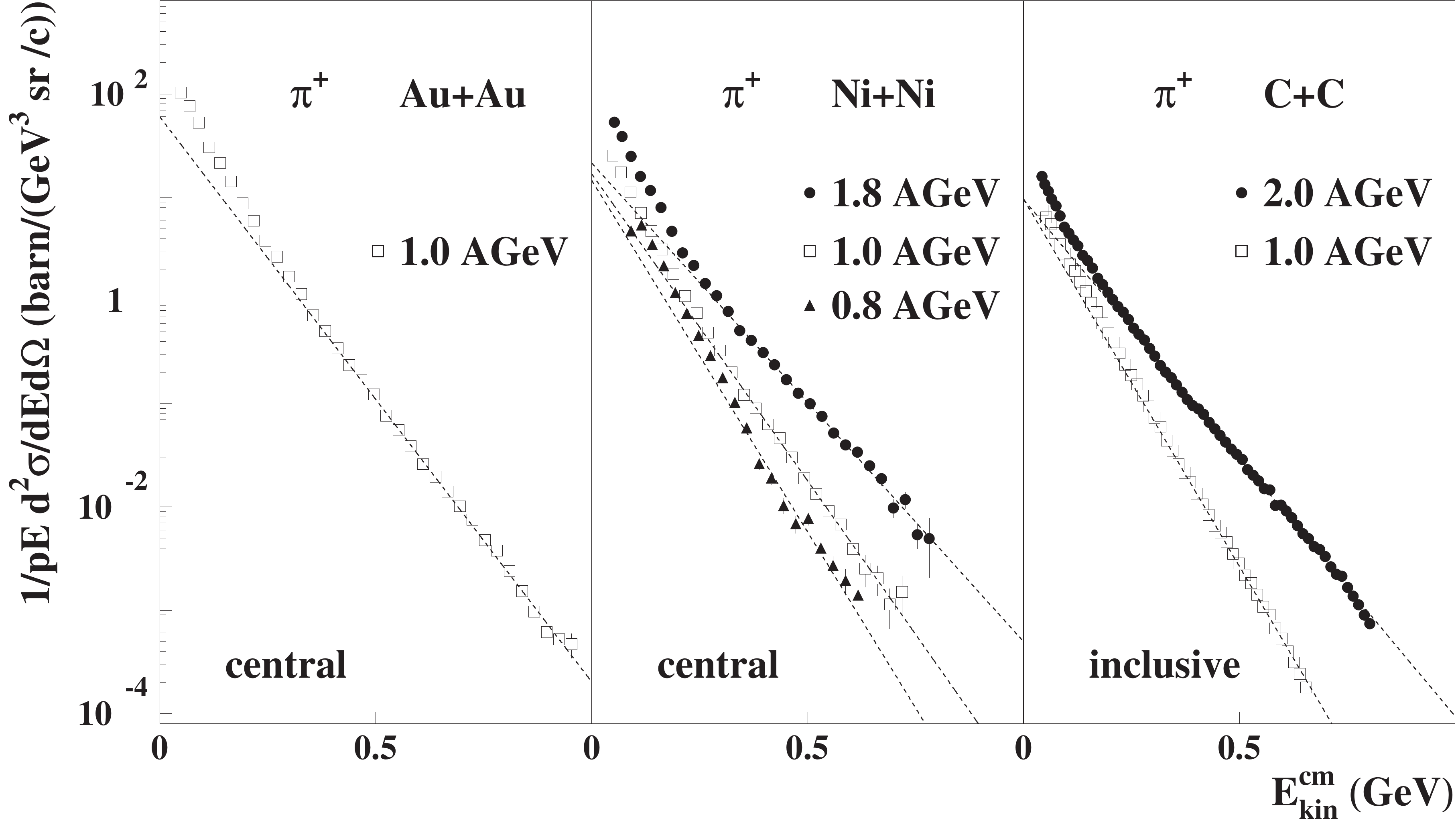,width=9cm}}
\centerline{\epsfig{file=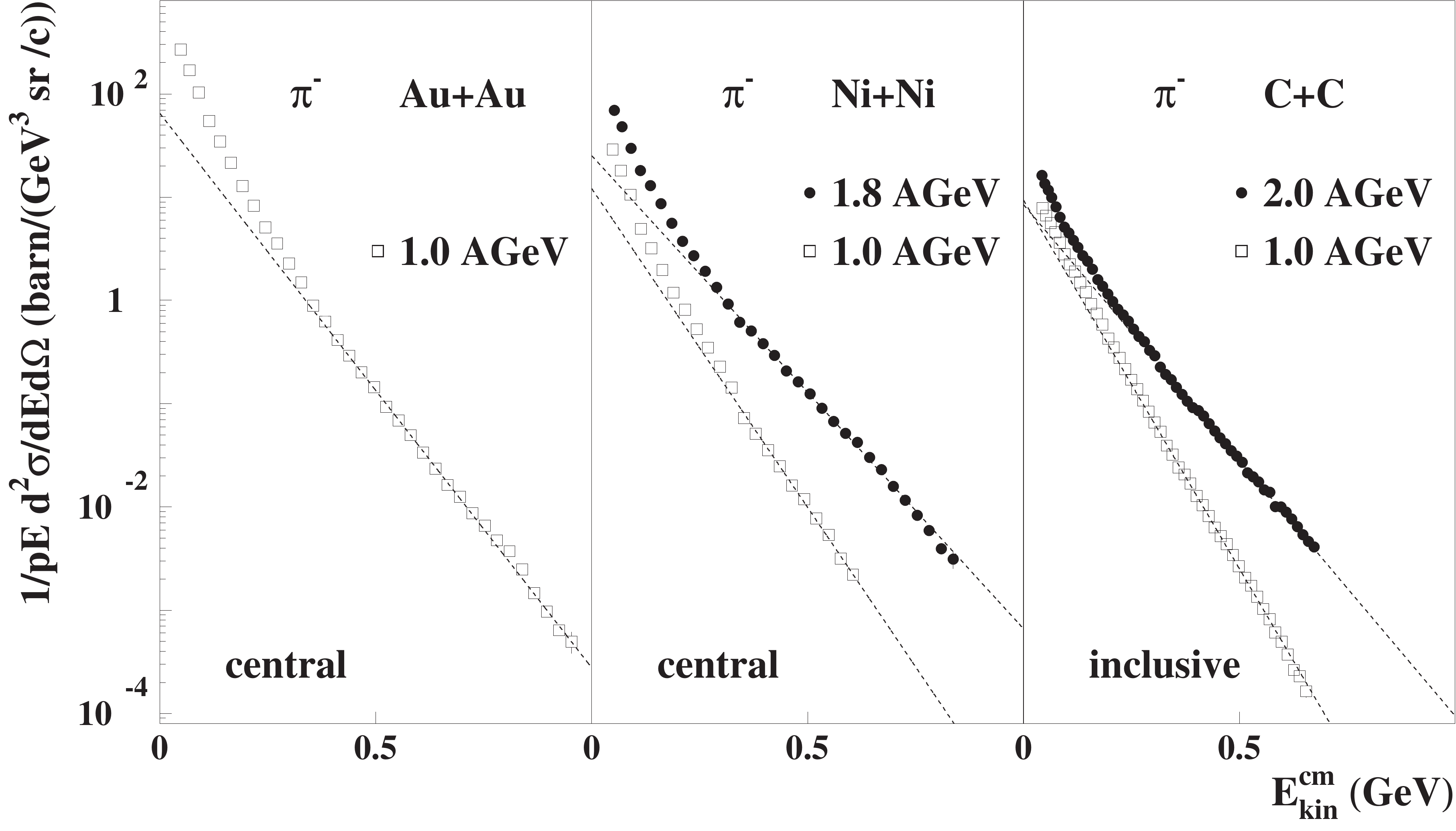,width=9cm}}
\caption{Spectra of positively (upper panel) and negatively (lower
panel) charged pions in the center-of-mass frame in a Boltzmann
representation for various reactions~\cite{Wagner_98,Sturm_diss}.}
\label{pi_p_m_all}
\end{figure}

The invariant cross sections for pion production in three
different systems are  shown in Figure~\ref{pi_p_m_all} as a
function of the center-of-mass kinetic energy. Above 0.4 GeV
kinetic energy in the c.m. system all spectra can be described  by
a Boltzmann distribution (exponential function in the given
presentation). The spectra also exhibit an enhancement at the
low-energy end if an exponential function is fitted in the
high-enery part. This excess is stronger for heavy systems. There
are different possible explanations for this enhancement. Late
time pion emission from $\Delta$-resonance
decay~\cite{Brockmann_84} and continuous pion
emission~\cite{Knoll_08} have been proposed. The most detailed
investigation to date has been done by Weinhold et
al.~\cite{Weinhold_98}.

\begin{figure}
\centerline{\epsfig{file=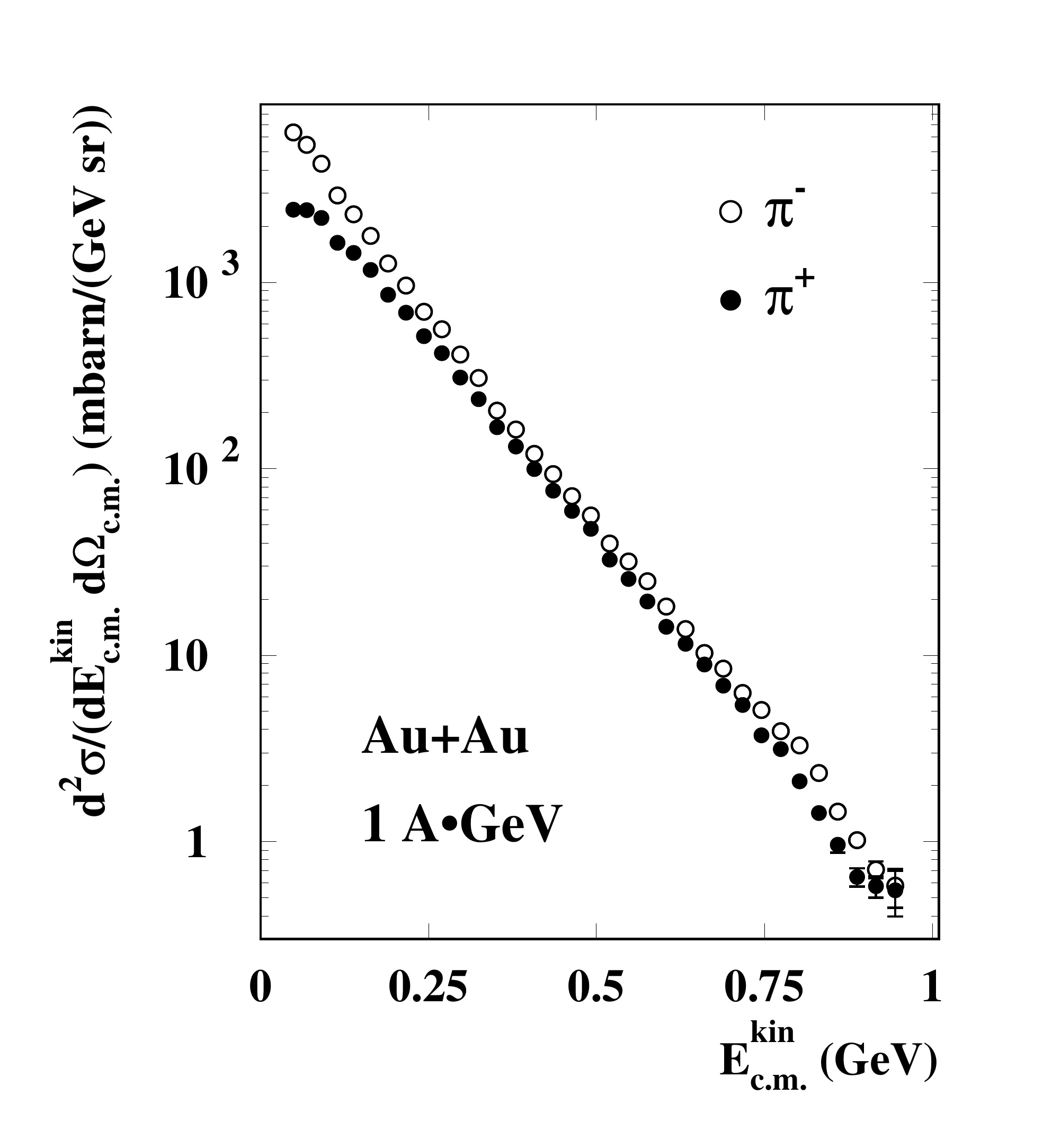,width=6cm}
\epsfig{file=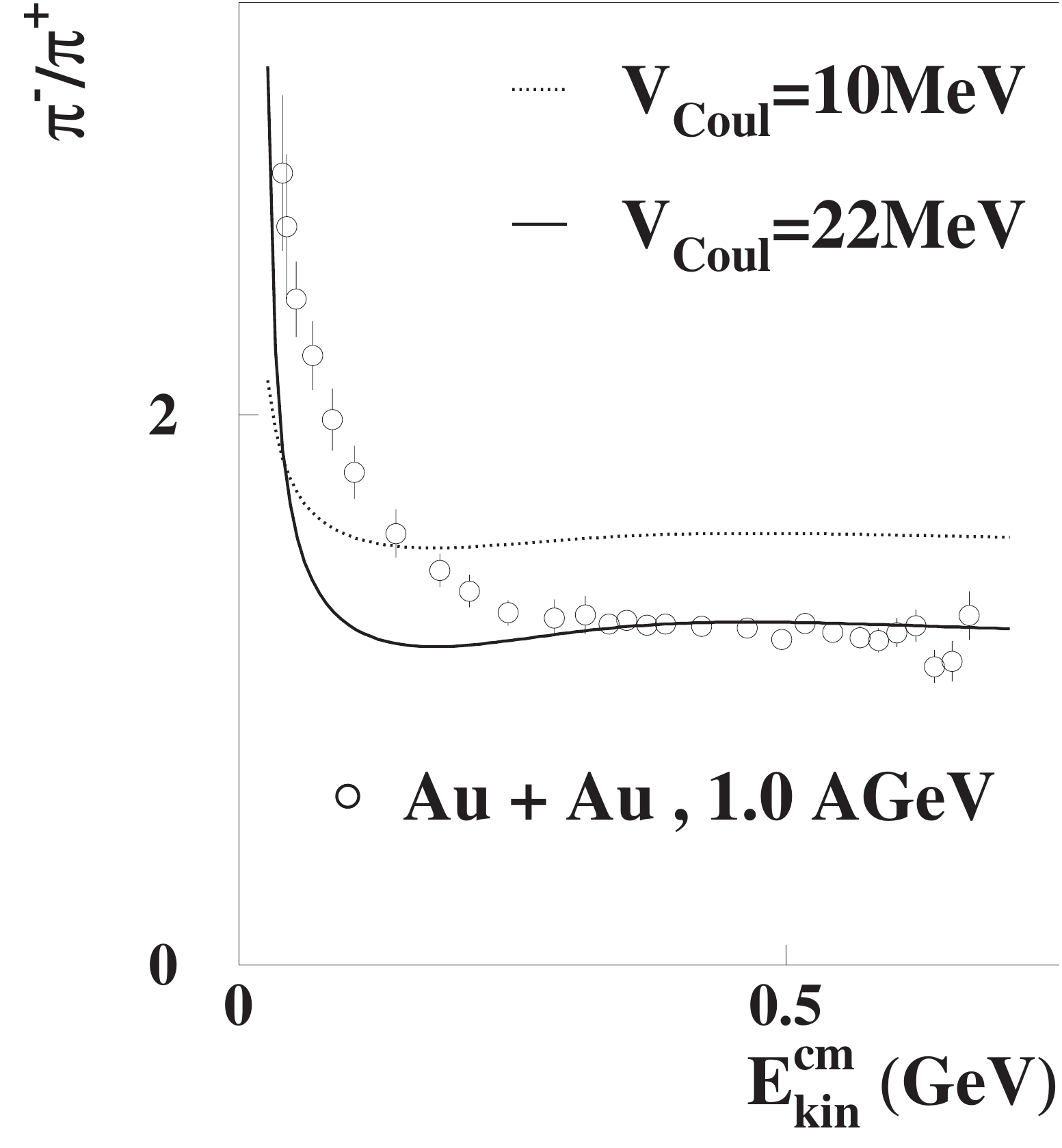,width=6cm}} \caption{Left
panel: Double differential cross sections of negatively and
positively  charged pions from central collisions of the reaction
Au+Au at an incident beam energy of 1 \AGeV and at an emission
angle of $ \theta_{\rm lab} = 44^\circ\pm4^\circ$ . Right panel:
Ratio of $\pi^-$/$\pi^+$ as a function of the kinetic energy of
pions for Au+Au collisions. From~\cite{Wagner_98}}
\label{pi_au_pm}
\end{figure}

As mentioned earlier, the ratio of the \pip and \pim
multiplicities reflect the N/Z ratio of the colliding nuclei and
the opposite Coulomb force. Figure~\ref{pi_au_pm} shows the double
differential cross sections of negatively  and positively charged
pions from central collisions of  Au+Au nuclei at an incident beam
energy of 1 $A$ GeV.
 The yield of \pim is higher due to the
neutron excess. The different shapes demonstrate the boost caused
by the Coulomb force. Similar observations have been made at the
Bevalac when studying forward emission of
pions~\cite{Sullivan_82}. The right panel of Figure~\ref{pi_au_pm}
compares the $\pi^-/\pi^+$ ratio as a function of the kinetic
energy of the pions with calculations using a static Coulomb
potential~\cite{Wagner_98} which is not able to describe the
dynamics of pion emission. For pions with lower kinetic energy a
weaker Coulomb force seems to act. This can be understood by the
time sequence of pion emission; high-energy pions are emitted
earlier than low-energy pions.

\begin{figure}
\centerline{\epsfig{file=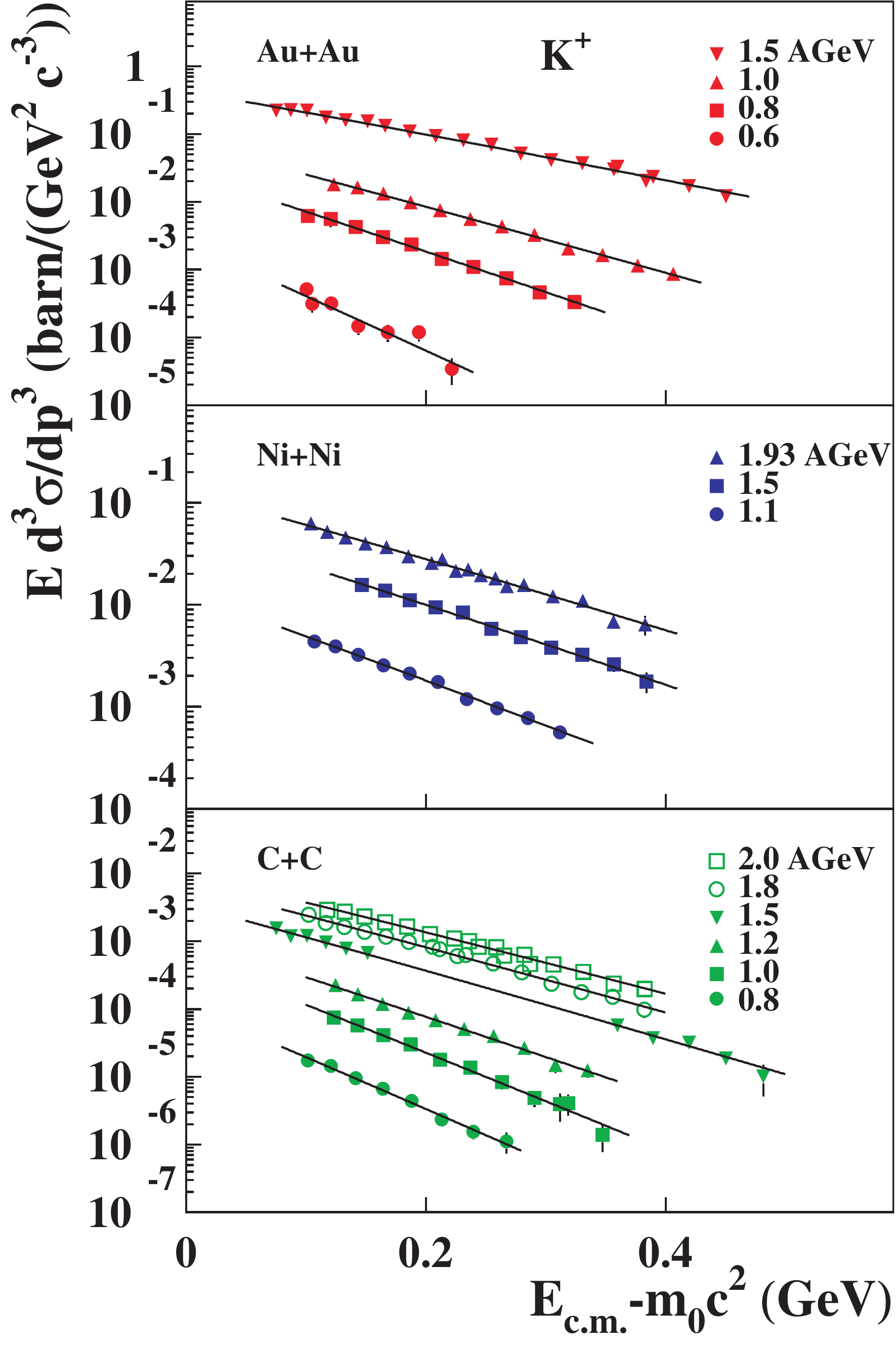,width=7.5cm}
\epsfig{file=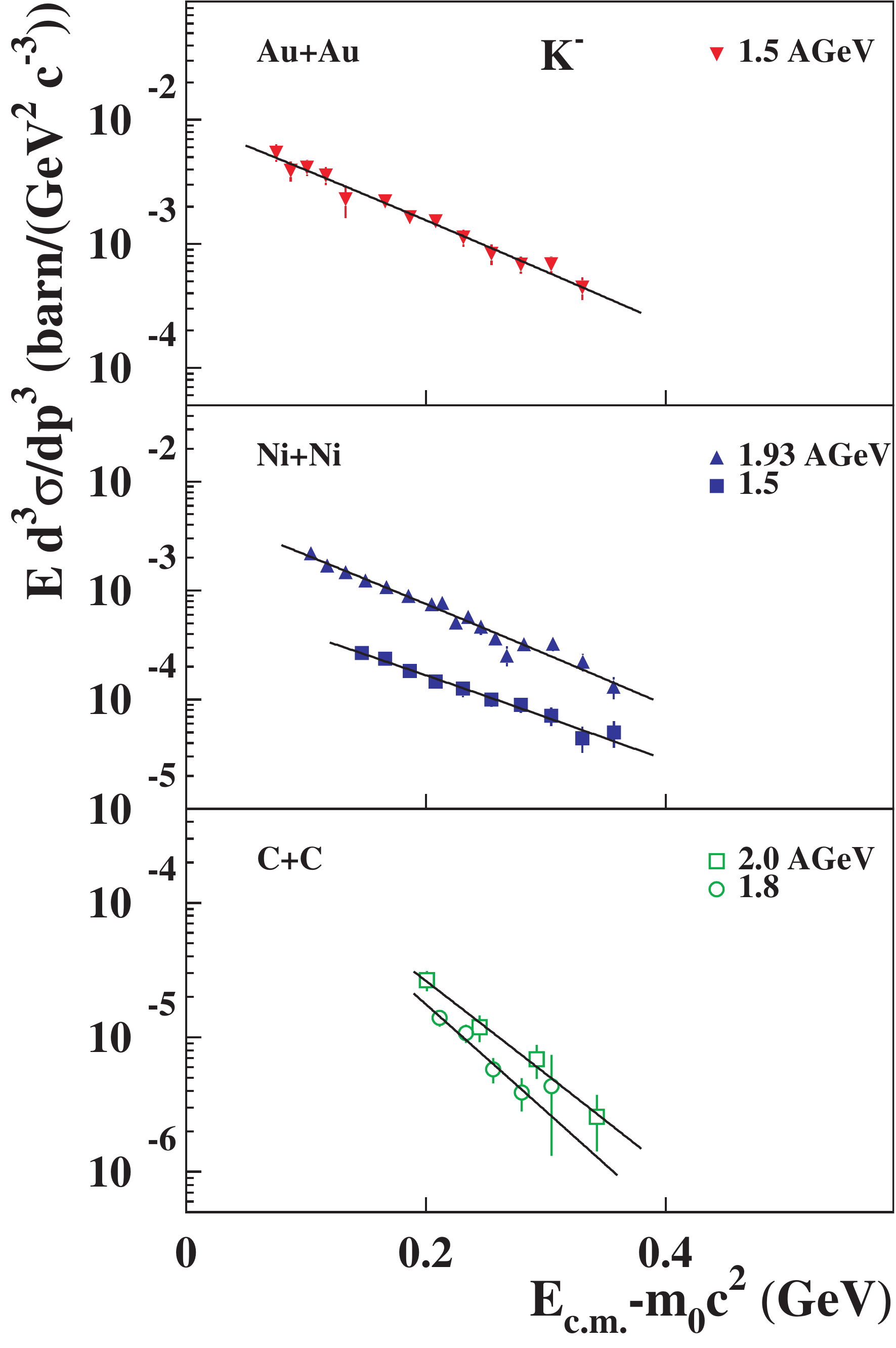,width=7.5cm}} \caption{
Inclusive invariant cross sections as a function of the kinetic
energy $E_{\rm c.m.} - m_{0}c^{2}$ for K$^+$ (left panel) and for
K$^-$ (right panel) for three systems and various beam energies at
mid-rapidity ($\theta_{\rm c.m.} = 90^\circ \pm 10^\circ$). From
\cite{KaoS}.} \label{midrap}
\end{figure}

Both K$^+$ and K$^-$ are produced in a complicated sequence of
interactions. Figure~\ref{midrap} shows K$^+$ and K$^-$ spectra at
mid-rapidity as a function of the kinetic energy $E_{\rm c.m.} -
m_{0}c^{2}$ for three different systems  and various beam
energies~\cite{KaoS}. The spectra have a Boltzmann shape to a very
good approximation.
The inverse slope parameters of the K$^+$ are always higher than
those of the K$^-$~\cite{Laue,Menzel,Wisn,KaoS}. Heavy systems
exhibit shallower inverse slopes than lighter ones.

\begin{figure}
\centerline{\epsfig{file=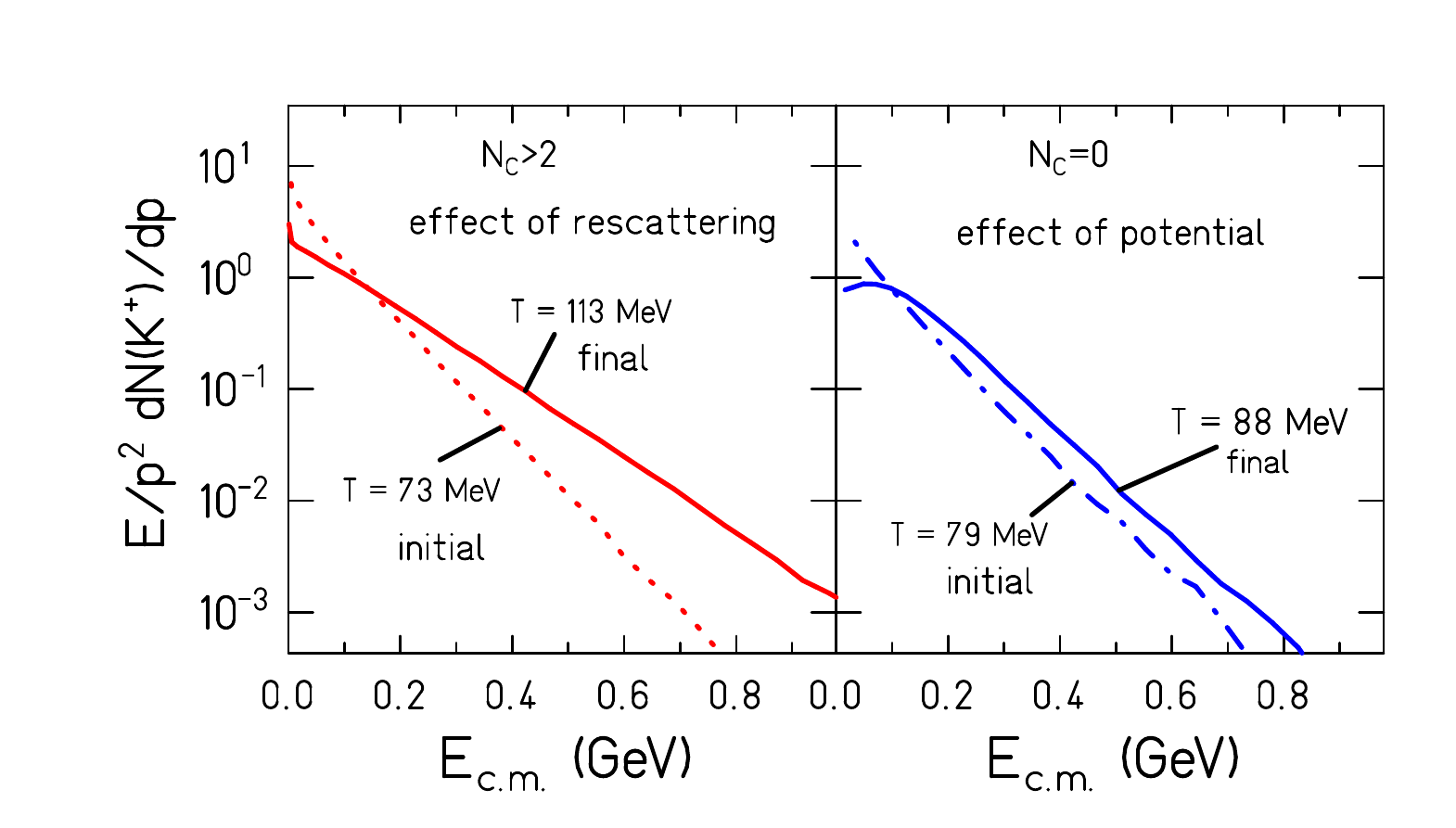,width=0.6\textwidth}}
\caption{ Influence of rescattering of the K$^+$ and of the
repulsive KN potential for central Au+Au collisions at 1.5
$A$~GeV.  Left panel: Initial and final distributions of kaons
which have scattered twice or more. Right panel: Influence of the
KN potential on the spectral shape demonstrated by selecting kaons
that never scattered ($N_C$ = 0) and comparing the initial and
final spectra~\cite{iqmd_PR}.} \label{spectra-effects}
\end{figure}

It is interesting to study the different effects contributing to
the K$^+$ slope in an IQMD study. Initially, the energy available
for K$^+$ production is low and the slope of the spectrum at
creation is therefore quite steep. Figure~\ref{spectra-effects}
shows the effect of re-scattering and of a possible kaon potential
on the final slope of K$^+$. Re-scattering makes the slope
shallower. This effect increases as the beam energy increases and
as the mass of the system increases as there is more scattering in
heavier systems. The  predicted repulsive K$^+$ potential would
increase the production threshold. When leaving the system, the
K$^+$ has to acquire its nominal mass and the excess energy serves
to accelerate it. This effect, however, is small and only visible
at low momenta. For experimental reasons low-momentum kaons  can
be studied best with neutral K$^0$ mesons. The corresponding
experiments are still ongoing~\cite{Merschmeyer_07,HADES_AS}.

\begin{figure}
\centerline{\epsfig{file=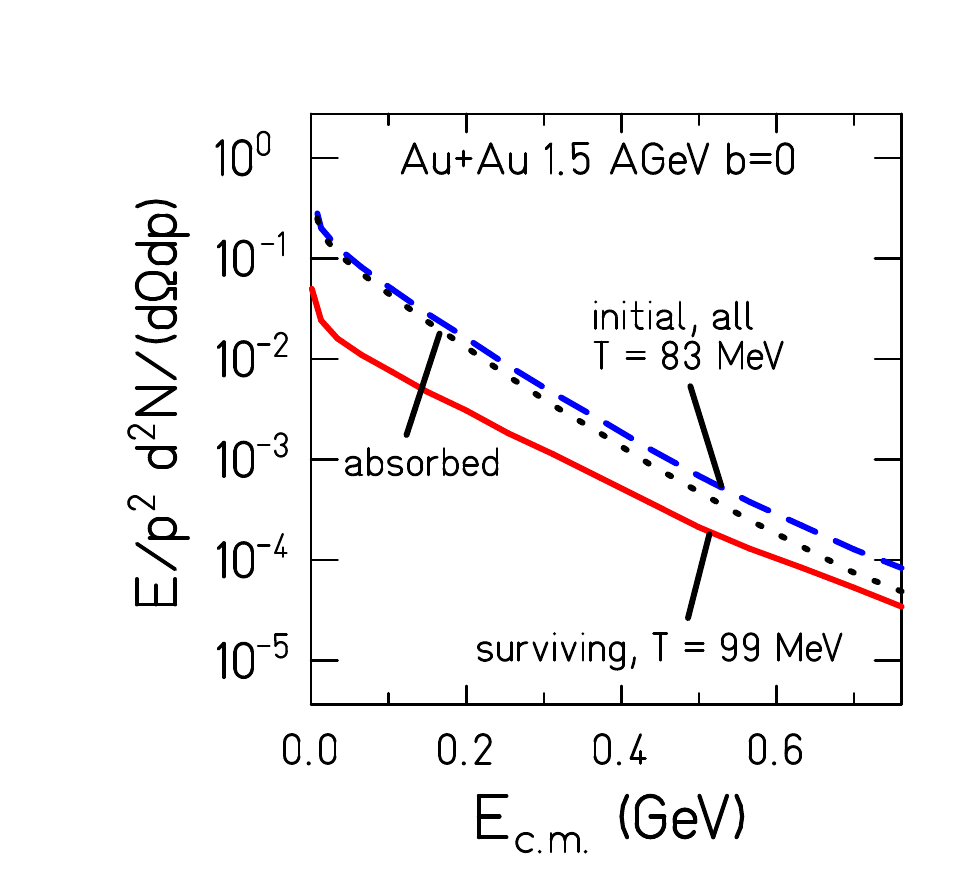,height=5.5cm}
\epsfig{file=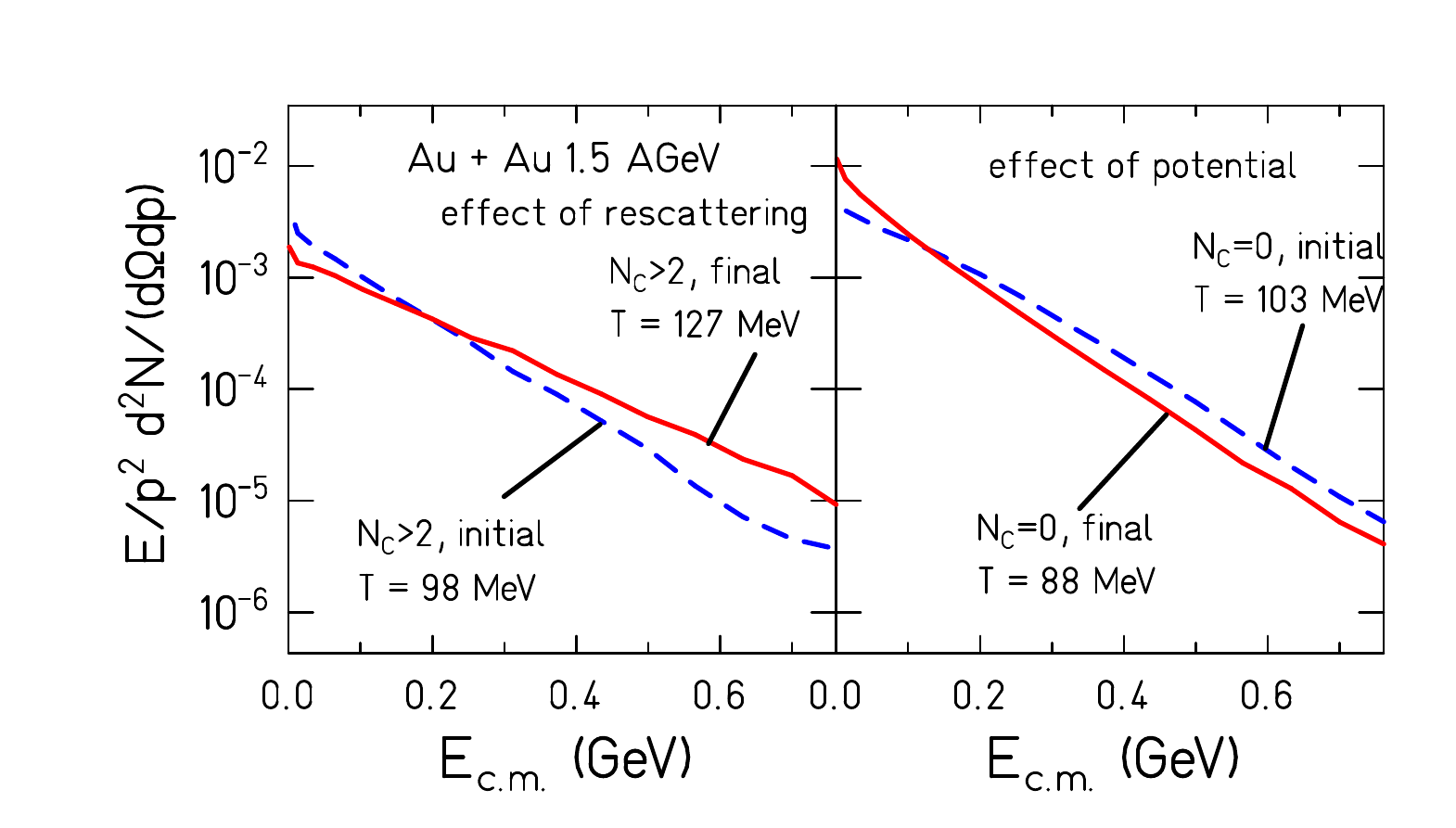,height=5.5cm}} \caption{
Left panel: Influence of absorption only demonstrated for
\mbox{Au+Au} at $1.5$~\AGeV based on IQMD calculations.  Right
panel: Effect of scattering of K$^-$ and of the attractive KN
potential~\cite{iqmd_PR}.} \label{spec_abs_nc_pot}
\end{figure}

Re-scattering has a strong influence on the K$^+$ spectra. The
spectral shape of the K$^-$ mesons is also influenced by
scattering. However, K$^-$ are mainly absorbed when interacting
with nuclei and absorption will change the slope in a
momentum-dependent way. The low-momentum K$^-$ undergo stronger
absorption which leads to an increase in the apparent temperature.
The various influences on the shape, resulting from re-scattering,
absorption, and the attractive KN potential are demonstrated in
Figure~\ref{spec_abs_nc_pot}.

\begin{figure}
\centerline{\epsfig{file=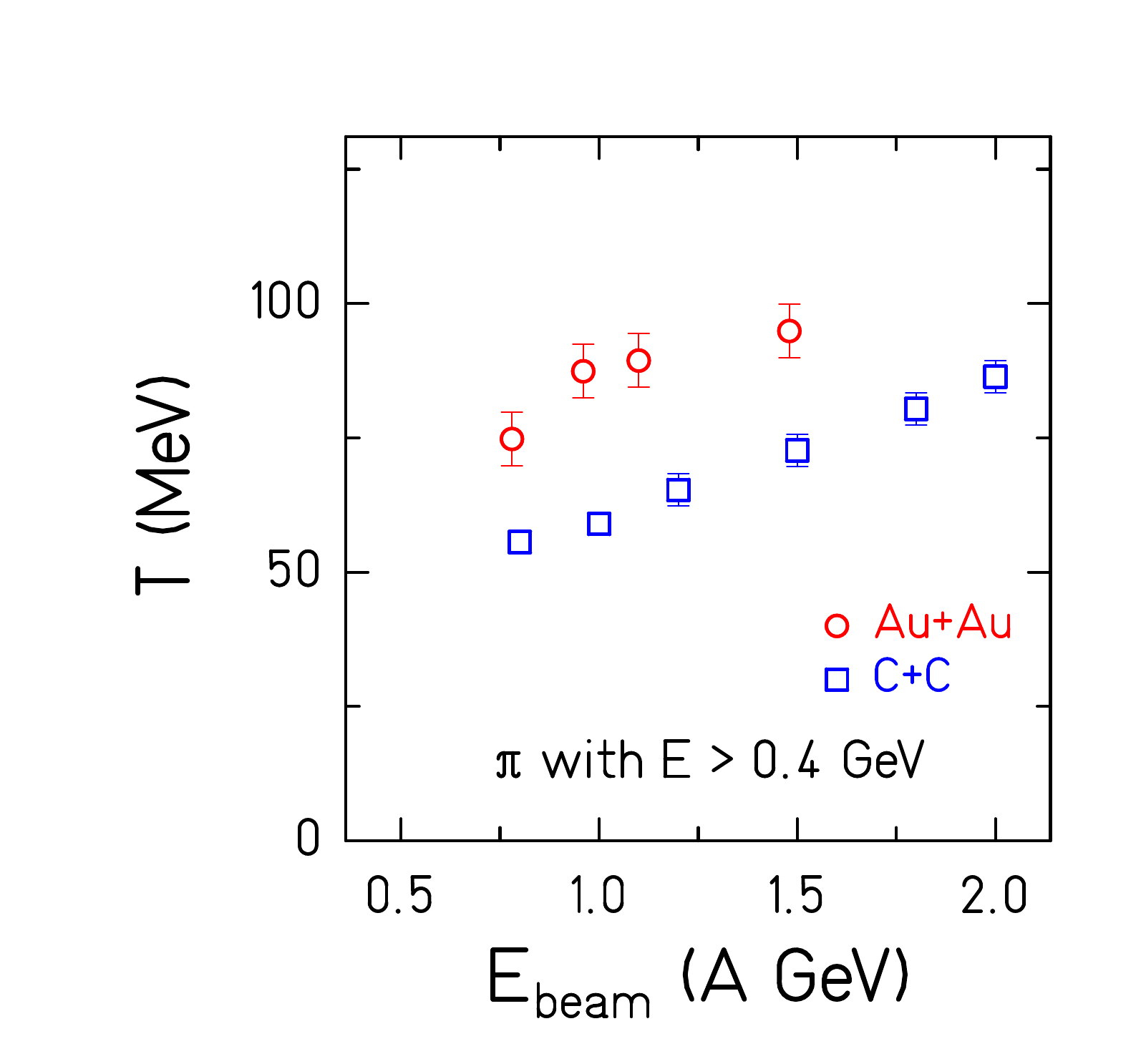,width=.48\textwidth}
\epsfig{file=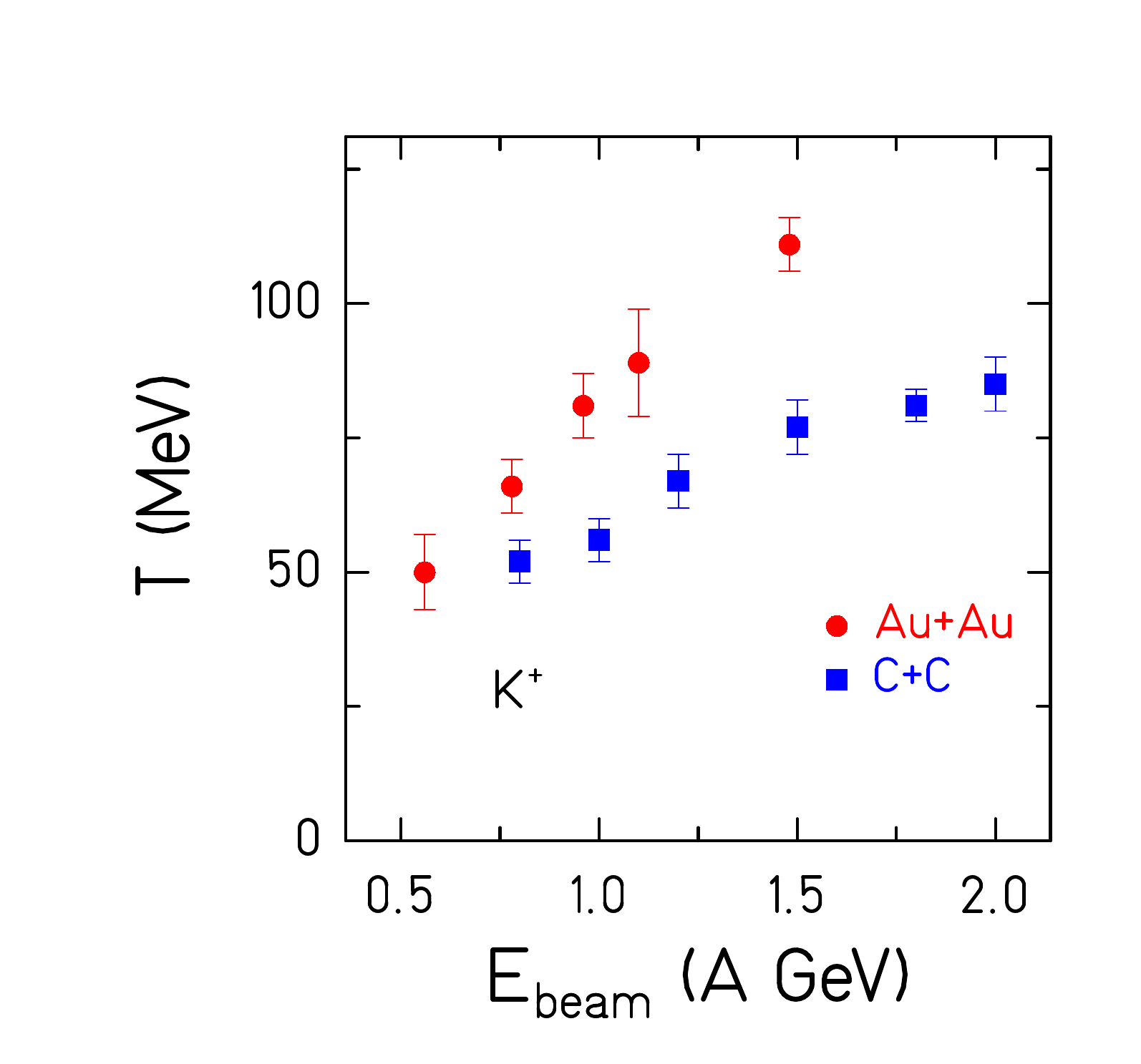,width=0.48\textwidth}}
\caption{Measured inverse slopes of high-energy pions (left panel) and of
K$^+$ mesons (right panel) from inclusive C+C and Au+Au collisions
at mid-rapidity~\cite{KaoS,sturm}.}
\label{tmidrap}
\end{figure}

Figure~\ref{tmidrap} gives a compilation of the measured inverse
slopes of inclusive $\pi^+$ (only high-energy part) and inclusive
K$^+$ for C+C and Au+Au collisions as a function of collision
energy, clearly showing the trends discussed before. The inverse
slopes increase with beam energy and at the same energy those of
the heavier systems are higher. It is interesting to note that the
values of the slopes of the high-energy pions and of the K$^+$
agree. They seem to reflect the temperature of the environment, if
this term can be used at all as a global equilibrium is not
achieved. Those of K$^-$ deviate as already shown in
Figure~\ref{midrap}.

\begin{center}
\begin{figure}
\epsfig{file=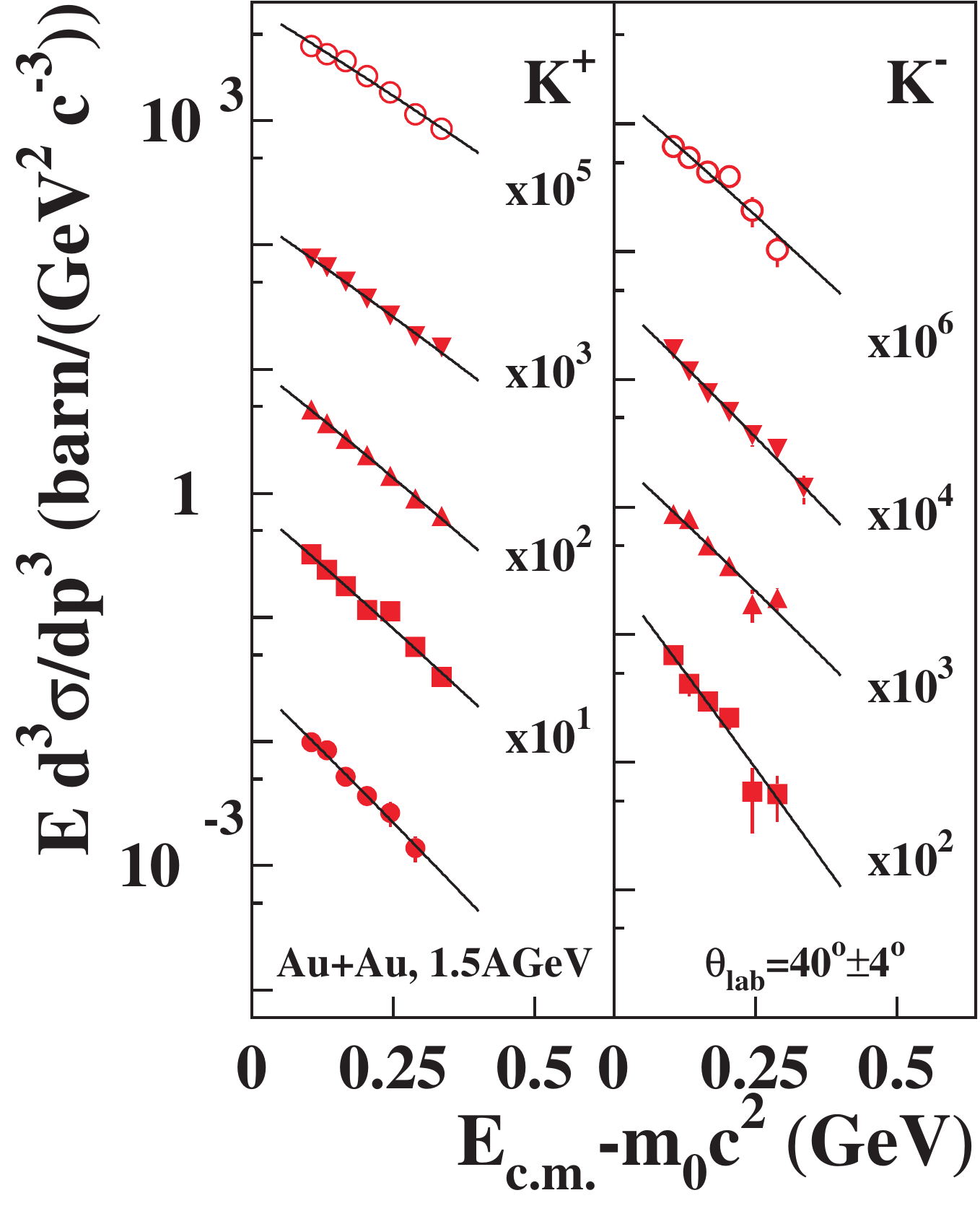,width=8cm}
\epsfig{file=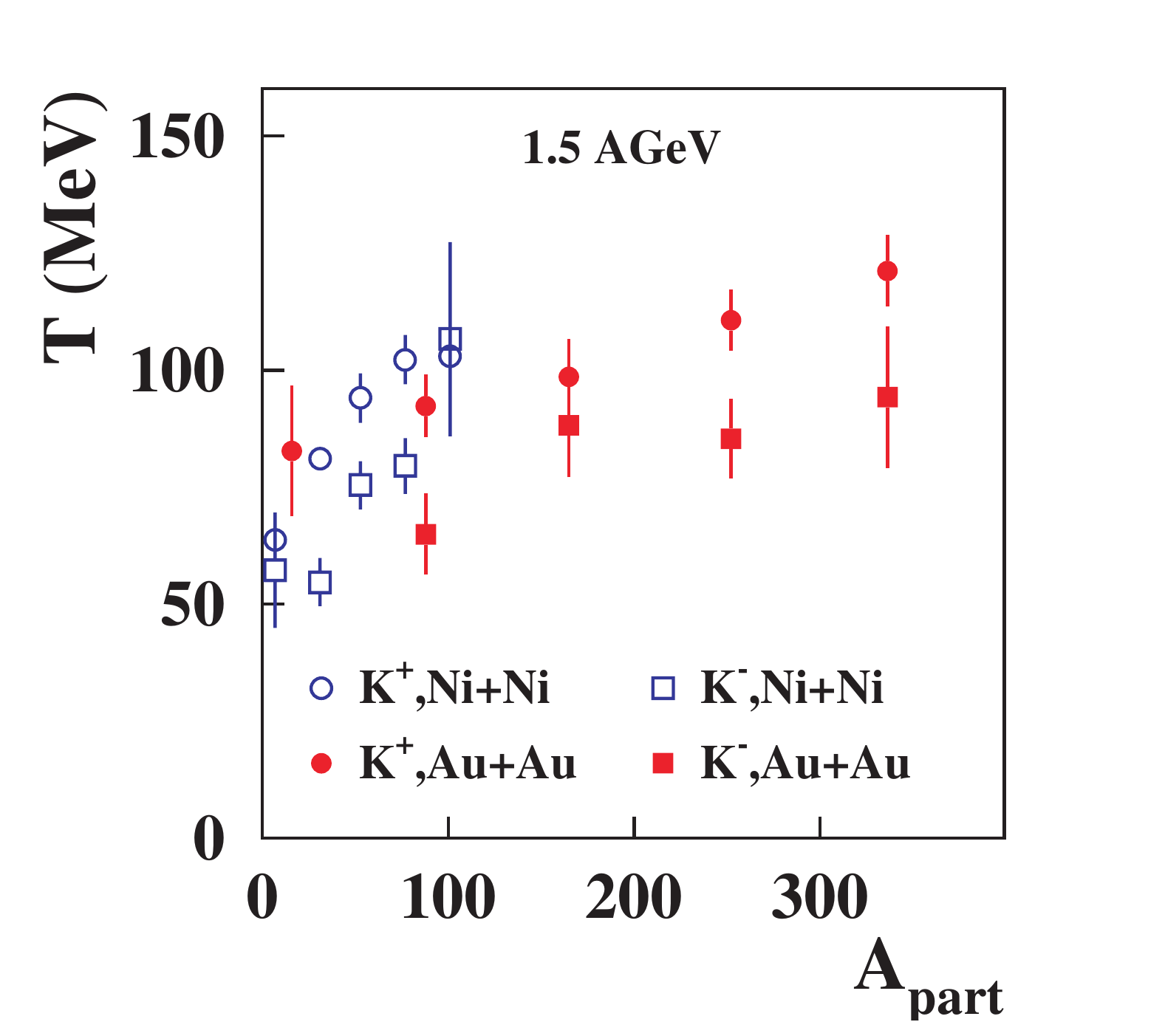,width=8cm} \caption{ Left
panel: Invariant cross sections for  \mbox{Au+Au} at $1.5$~\AGeV
close to mid-rapidity for different centralities. The uppermost
spectra correspond to the most central collisions. The subsequent
bins are shown from top to bottom with decreasing centrality. The
lines represent fits to the KaoS data~\cite{KaoS}. Right panel:
Summary of inverse slope parameters~\cite{KaoS}. } \label{spec_cm}
\end{figure}
\end{center}

Finally, we investigate the impact parameter dependence of the
kaon spectra. Figure~\ref{spec_cm} (left panel)  shows  the
spectra measured by the KaoS Collaboration for Au+Au reactions at
$1.5$~\AGeV for different centralities for K$^+$ and K$^-$.  The
inverse slope parameters, $T$, are evaluated  as a function of
centrality or $A_{part}$. This demonstrates that the slope
parameters are weakly depending on centrality and that the slope
of K$^-$ is always about 20 MeV lower than that of K$^+$.

\subsection{Conclusions}

Heavy ion collisions at 1 to 2 \AGeV incident energy probe nuclear
matter at about two to three times its normal density. The
production and emission of \kp and of \km exhibits distinct
differences. The key points are summarized in a model study
showing the emission time of the two kaon species and the
corresponding density profiles of the medium within the IQMD
approach~\cite{iqmd_PR}.
\begin{figure}[]
\centerline{\epsfig{file=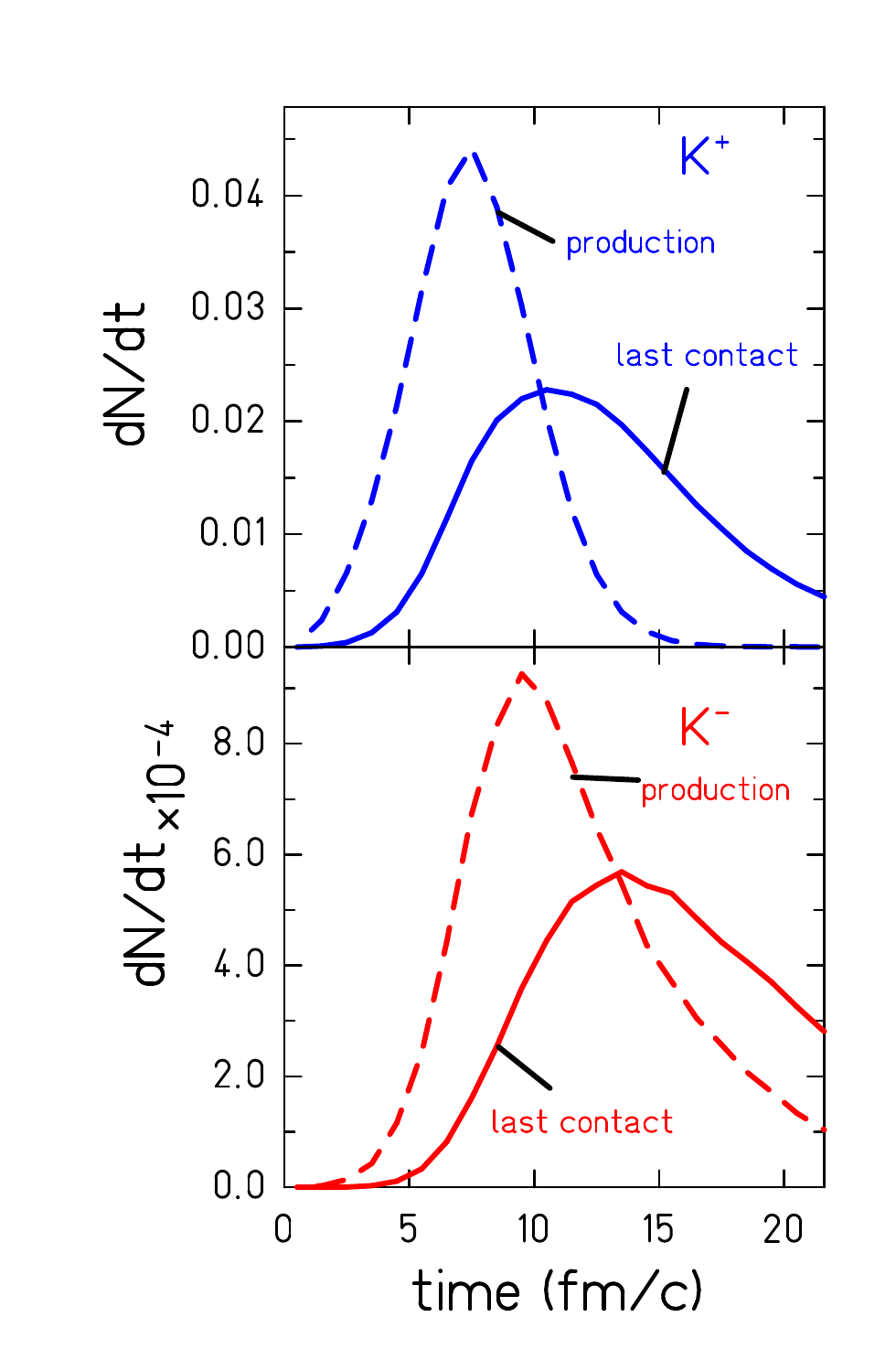,width=5cm}
\epsfig{file=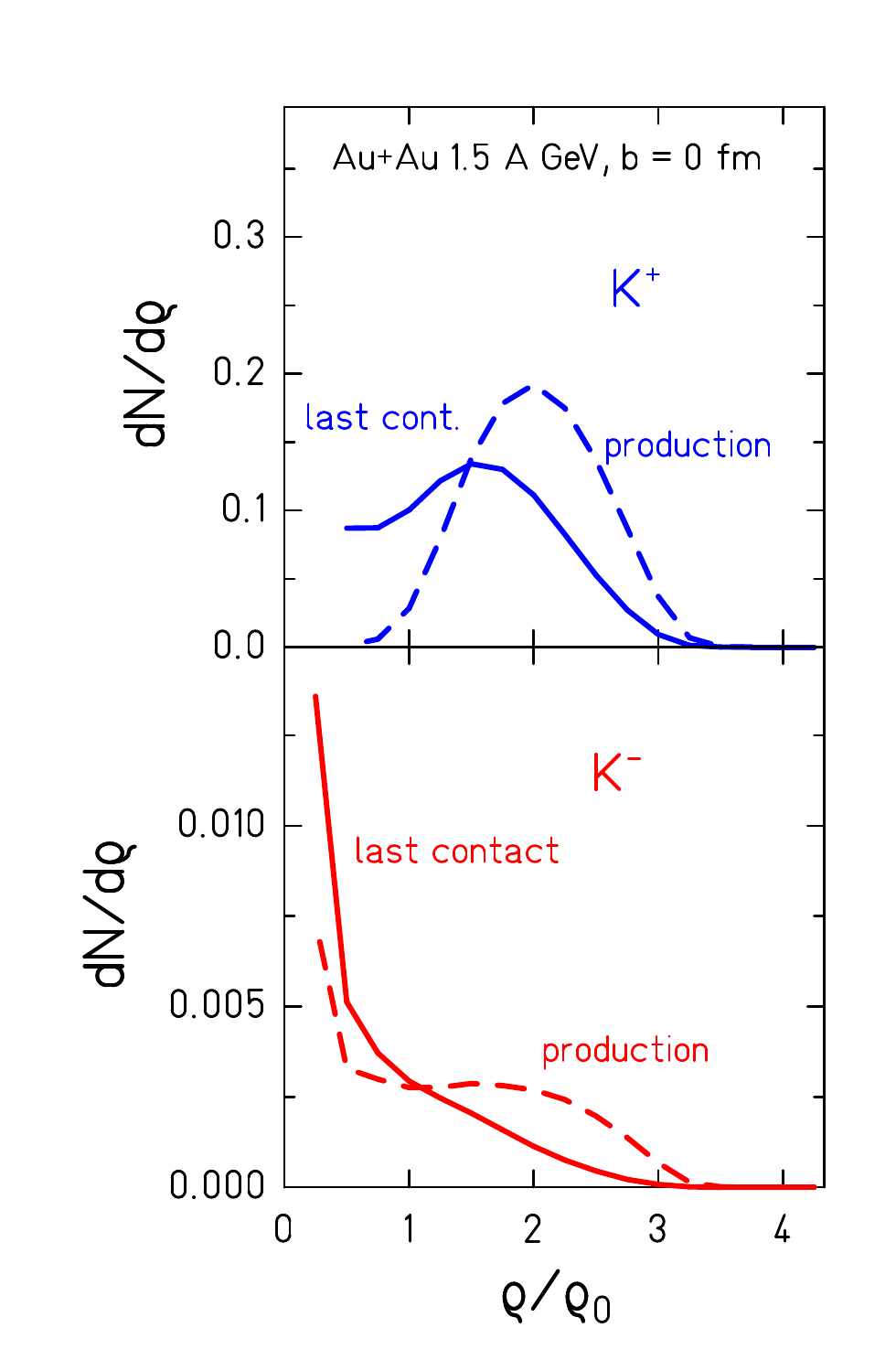,width=5cm}}
 \caption{ Left panel: Time profiles for production (dashed lines) and
 last contact (solid lines) of K$^+$ (top) and K$^-$ (bottom)
 mesons. Only those \km are shown which finally leave the system.
 Right panel: Density distribution at the point of production (dashed lines)
 and at the point of last contact (solid lines) for K$^+$ (top) and K$^-$ (bottom) mesons.
 The simulations are for central Au+Au collisions at 1.5 $A$ GeV. Figures are taken from \cite{iqmd_PR}.}
\label{density-time}
\end{figure}
The left panel of Figure~\ref{density-time} shows  the time
distributions at production and at last interaction. K$^-$ are
produced later than K$^+$ and they also leave the system later.
The right panel of Figure~\ref{density-time} shows the density of
the medium at production and at last contact. The bulk of the
K$^+$ is produced when the density is twice nuclear matter
density. \kp are emitted predominantly at a density of $\rho$ =
$1.5$ $\rho_0$. Their yield cannot be changed between production
and emission due to strangeness conservation. Therefore, the \kp
yield carries information from the high-density phase.  As \kp
production is at or below the NN threshold, its yield is very
sensitive to the dynamics in the fireball. \kp production serves
as an ideal tool to extract information on the nuclear EOS. In
order to avoid uncertainties both experimentally and
theoretically, it is useful to study ratios of \kp
multiplicities~\cite{sturm}, e.g.~the ratio of multiplicities from
Au+Au and C+C collisions.
 From this ratio a rather soft EOS
with a compression modulus of around 200 MeV has been
extracted~\cite{fuchs,iqmd_eos}. This is  in agreement with
results from flow studies~\cite{Aich_Rev,ritter97}.

The observed K$^-$ mesons are predominantly produced and emitted
late and from a region of density below nuclear matter density,
whereas the K$^+$ are produced at twice nuclear matter density. It
is obvious that heavy ion reactions in this energy range are well
suited to study K$^+$ potential effects. This is not the case for
K$^-$ production, since the effects are expected to be small.
Other observables, like the azimuthal emission pattern, also show
sensitivity to in-medium interactions~\cite{Shin,Uhlig}.


\section{Hadron Production from AGS to RHIC}

In this section we discuss the evolution of spectra of identified
hadrons as the center-of-mass energy is increased from AGS to
RHIC. For AGS and SPS we will summarize the spectra in terms of
inverse slope parameters, $T$, and show how  radial flow develops
as a function of energy. At RHIC energies we will use the wealth
of precise data on radial and elliptic flow to argue that
collectivity develops at the partonic level.

We start with a discussion of the transverse momentum spectra and
the way they are analyzed and important dynamical parameters are
extracted using the example of spectra from RHIC experiments. The
methods developed are applicable to all energies discussed here.

\begin{figure}[]
\centerline{\psfig{file=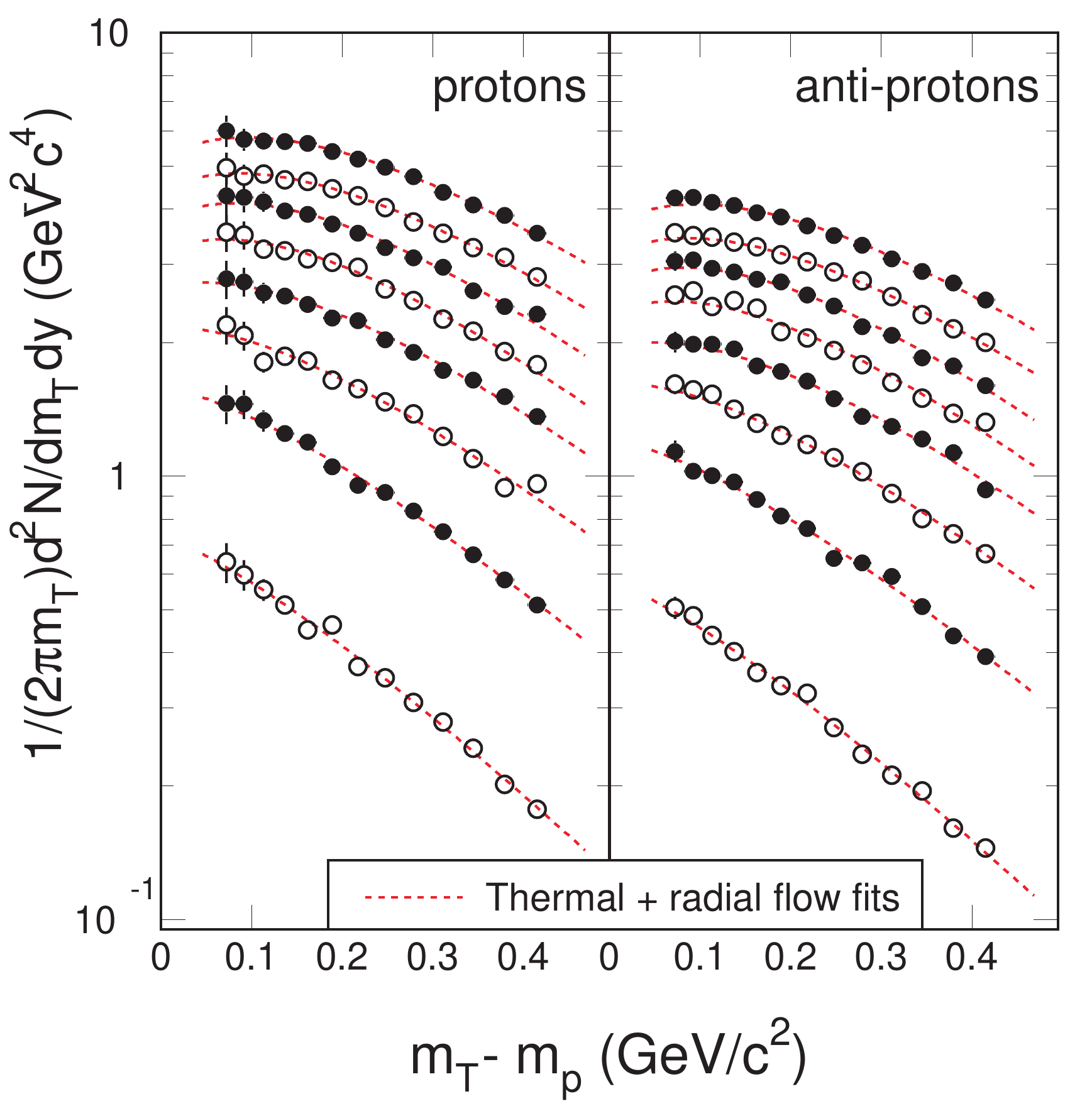,height=8.50cm}}
\caption{Mid-rapidity ($|y|\le0.5$) proton (left column) and
antiproton (right column) transverse mass distributions from Au+Au
collisions at \rts = 130 GeV. From bottom to top the curves are
ordered according to centrality bins, 70-80\%, 60-70\%, 50-60\%,
40-50\%, 30-40\%, 20-30\%, 10-20\%, and 0-10\%.  Results from
thermal model fits are shown as dashed lines.} \label{nu_fig5}
\end{figure}

Figure~\ref{nu_fig5} shows mid-rapidity proton and antiproton
transverse mass distributions  for several centrality bins. The
data are from Au+Au collisions at \rts = 130 GeV \cite{starp130}.
The transverse mass, $m_T$, is given by $m_T = \sqrt{(p_T^2 +
m_0^2)}$, with $m_0$ the rest mass of the proton (antiproton). It
is evident that the distributions become more convex as centrality
increases. The change in shape is an indication for strong radial
flow.  In the presence of collective flow the transverse mass
distributions, especially for heavy mass particles, will not have
a simple exponential shape at low transverse
mass~\cite{Siemens_79}. This effect becomes particularly strong
when the temperature is low compared to the collective velocity.
Two-parameter fits~\cite{na44flow}  of the freeze-out temperature,
$T_{fo}$, and the collective velocity, $\beta_t$, motivated by
hydrodynamics, are useful to separate the collective motion from
random thermal motion in the measured spectra. Here we apply blast
wave fits from Reference~\cite{heinz93}. These fits simultaneously
describe experimental spectra of charged pions~\cite{starpion},
kaons~\cite{starkaon}, protons and antiprotons. A velocity profile
$\beta_t(r) = \beta_s (r/R)^{0.5}$ is used in the fit, where $R$
and $\beta_s$ are the radius and the surface velocity of the
source, respectively. The fit-results are shown as dashed lines in
Figure~\ref{nu_fig5}. The collective velocity parameter $< \beta_t
>$  increases from about 0.41 $c$ to 0.55 $c$ for the most  central
collisions.

\begin{figure}[t]
{\psfig{file=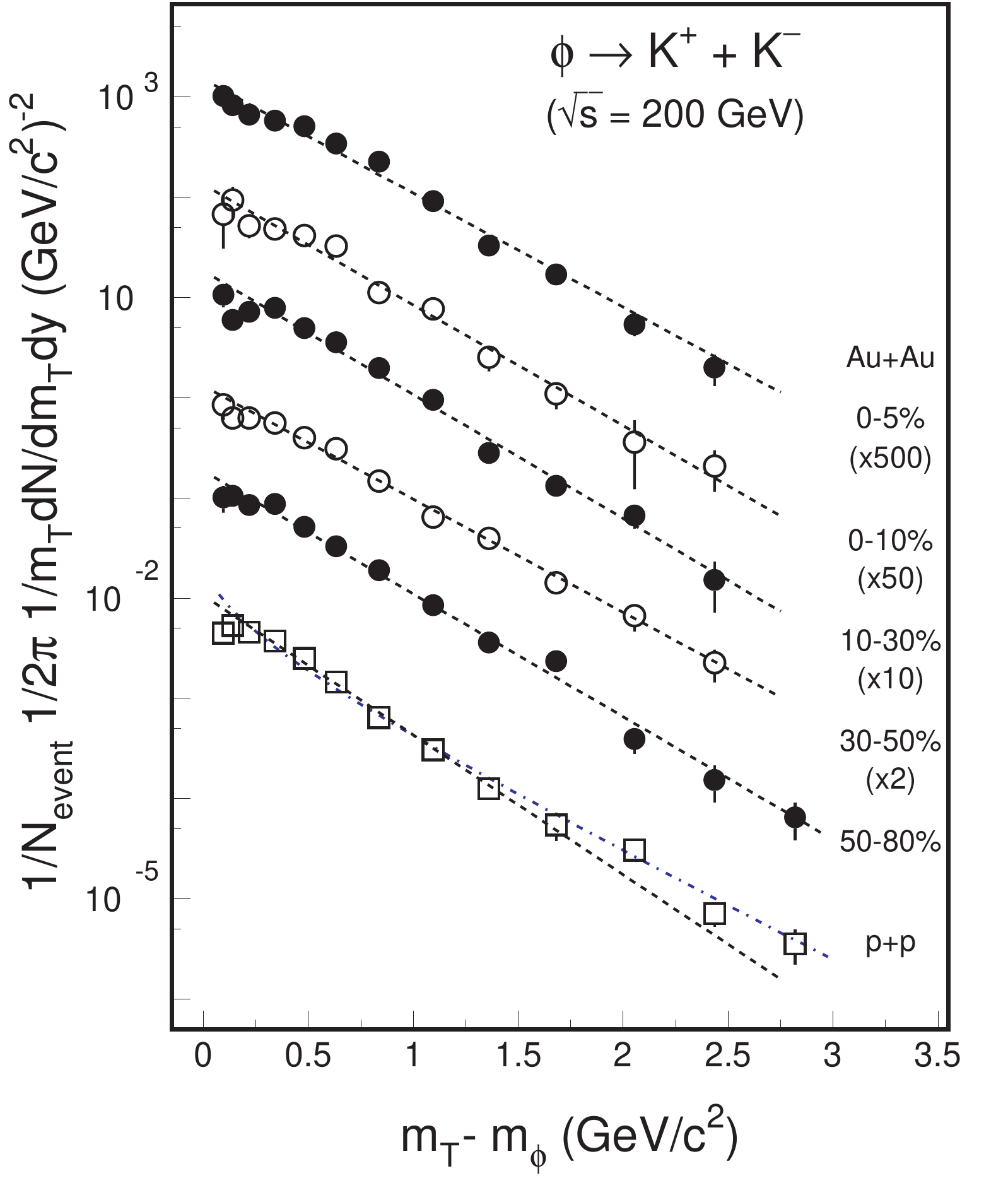,height=9cm}}
{\psfig{file=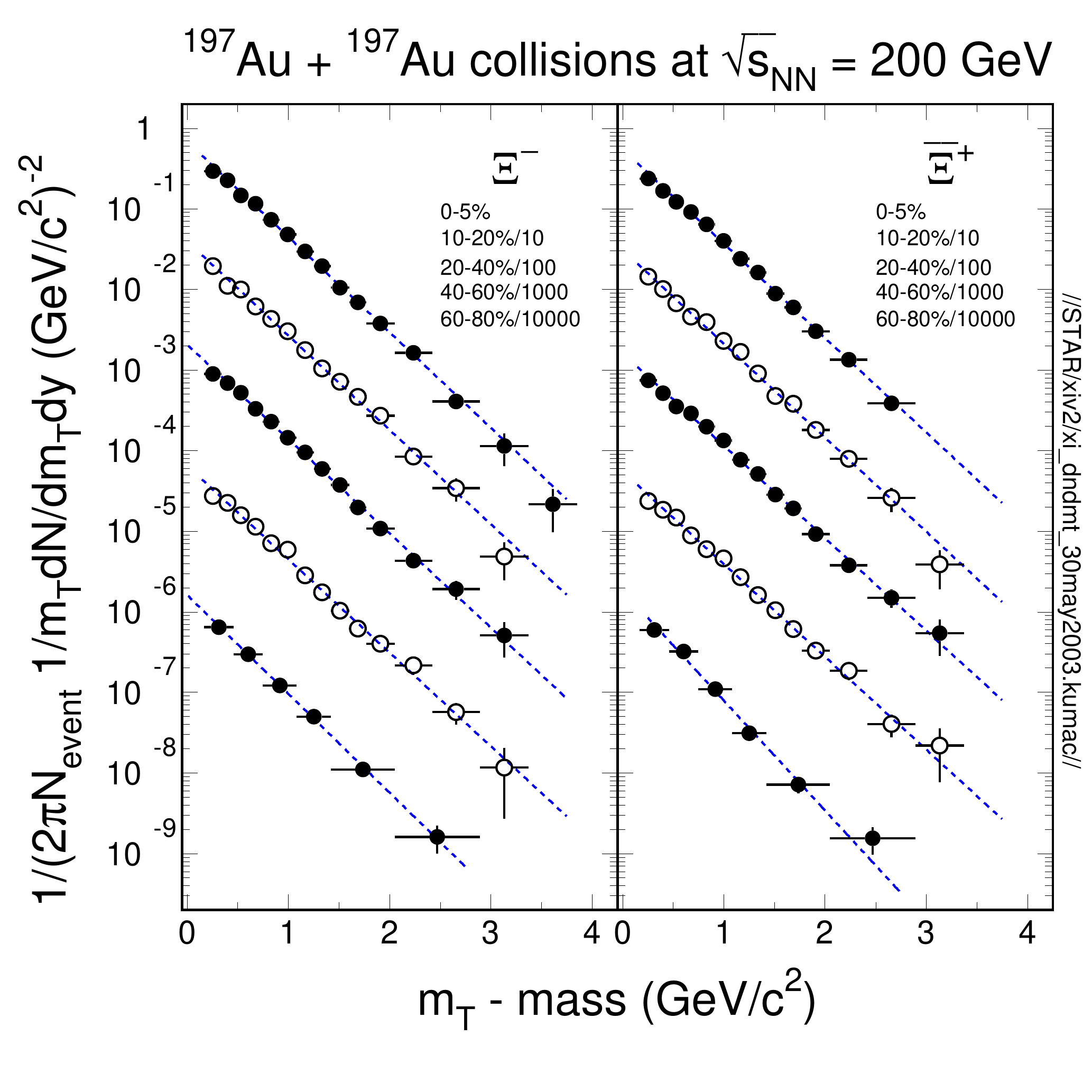,height=9cm}}
\caption{Mid-rapidity transverse mass  distributions from Au+Au
and p+p collisions at \rts =  200 GeV for $\phi$ mesons (left
panel). Dashed lines represent exponential fits to the data and
the dot-dashed line is the power-law fit to the p+p spectrum. In
the right panel $\Xi$ particles are shown. Dashed lines represent
exponential fits. The data are from STAR~\cite{fig22_ref}.}
\label{nu_fig7}
\end{figure}

Figure~\ref{nu_fig7} shows $\phi$ meson and $\Xi$ hyperon
transverse mass distributions from 200 GeV Au+Au and p+p
collisions. The spectral shapes are quite different from the ones
shown in Figure~\ref{nu_fig5}. All spectra with the exception of
the $\phi$ mesons emitted in p+p collisions have an exponential
form:
\begin{equation}
f_{exp} = A \cdot e^{-m_T/T},
\end{equation}
where $A$ and $T$ are the normalization constant and inverse slope
parameter, respectively. The dashed lines in the figure represent exponential fits.
In the case of $\phi$ mesons emitted in p+p collisions, a
power-law fit gives better results:
 \begin{equation}
f_{p} = A \cdot (1+{p_T}/{p_0})^{-n}.
\end{equation}
Here $A$ is the normalization constant and $p_0$ and $n$ are free
parameters that describe the shape of the distribution. The
dash-dotted line represents the fit. In general, mesons emitted from p+p collisions
are following a power-law type distribution with a $p_0$ of
about 12 GeV/c and a $n$ of six, while the baryons are closer to
an exponential function~\cite{ua196}.

\subsection{Systematics of Spectra and Slopes at AGS and SPS}

The AGS program has yielded a large body of emission spectra
mainly for pions, kaons, and protons.  At the SPS there is an even
larger wealth of data. Rather than reviewing the data in detail,
we will concentrate on summarizing the emission spectra in terms
of inverse slope parameters or apparent temperatures.


\begin{figure}
\centerline{\epsfig{file=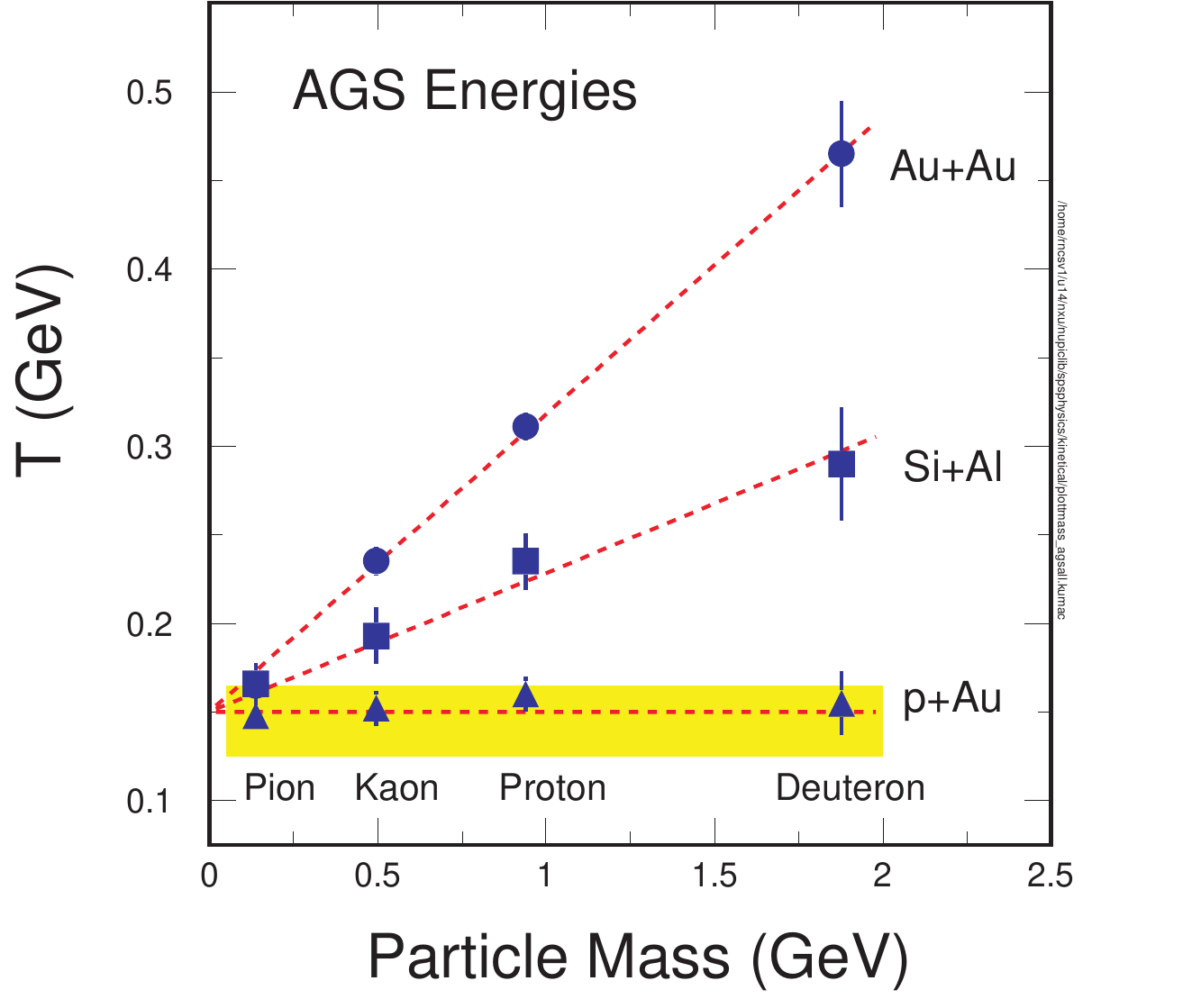,height=8.0cm}}
  \caption{Inverse slope parameter, $T$, of pions, kaons, protons,
  and deuterons from central collisions at the AGS (\rts $ \sim 5$ GeV) for
   \pau, \sial, \auau.  }
\label{nu_fig1}
\end{figure}


\begin{figure}
\centerline{\epsfig{file=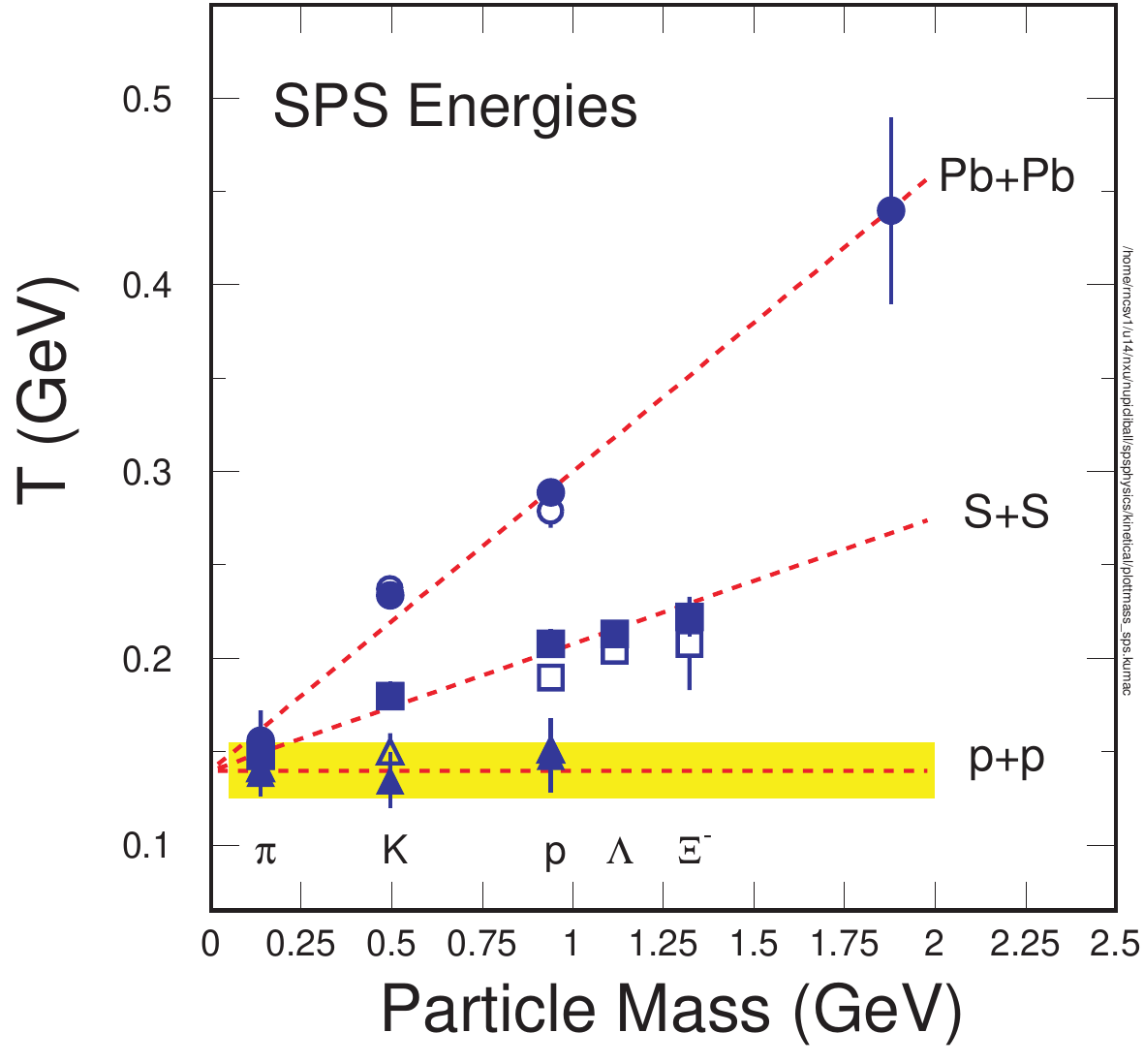,height=8.0cm}}
\caption{Inverse slope parameter, $T$,  of pions, kaons, protons,
$\Lambda$, and $\Xi$ from central S+S and Pb+Pb  collisions and
minimum bias p+p collisions at the SPS (\rts $\sim 17-20$ GeV).}
\label{nu_fig2}
\end{figure}

Figure~\ref{nu_fig1} shows the systematics of the extracted
inverse slope parameters as a function of the mass of the emitted
particles at AGS energies for three systems with different total
mass. The data are taken from the E802 experiment~\cite{E802}.
Figure~\ref{nu_fig2} gives the corresponding systematics for SPS
energies. Here the data are taken from  the NA44
experiment~\cite{na44flow}. At both energies a clear trend
emerges: a linear relationship between inverse slope parameter and
particle mass. This dependence can be parameterized as a
freeze-out temperature plus a linear term in particle mass with a
coefficient that is related to the flow velocity. The p+p and the
p+nucleus systems do not show any dependence on mass. Thus, in
this approach, they do not show collective expansion. For the
heavier systems we observe a collective expansion velocity that
increases with the mass of the system. A freeze-out temperature of
140 to 150 MeV has been extracted. The systematics observed in
Figure~\ref{nu_fig1}  and in Figure~\ref{nu_fig2} has been taken
as a strong indication of the existence of collective radial
expansion in heavy ion collisions in this energy
range~\cite{xu_NA44,nxu01}.


\begin{figure}
\centerline{\psfig{file=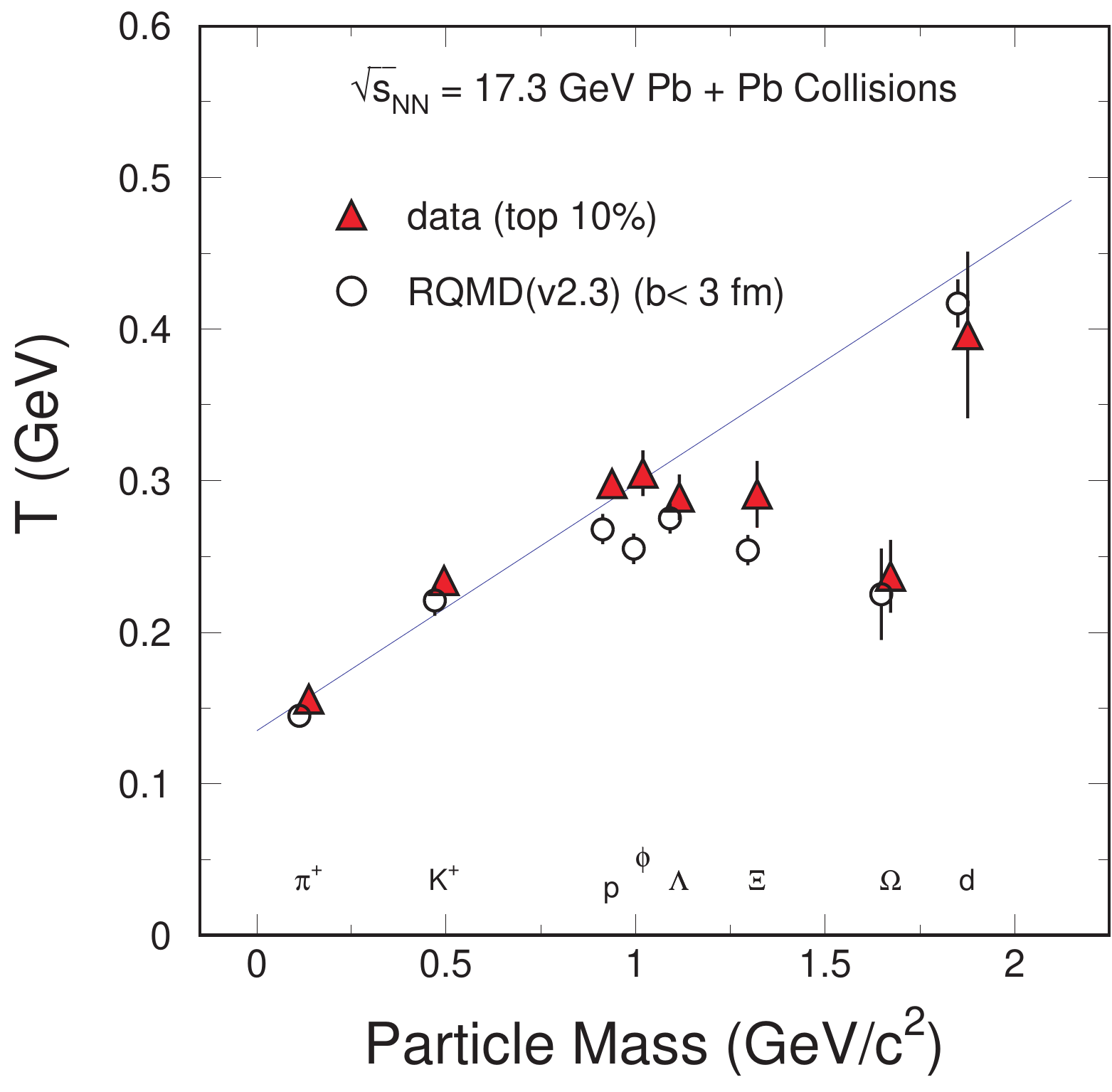,height=9.0cm}}
\caption{Measured inverse slope parameter, $T$, (filled triangles)
and RQMD model predictions (open circles) as a function of
particle mass of central  Pb+Pb collisions at  \rts  $\sim 17.3$
GeV.  The line serves to guide the eye. } \label{nu_fig3}
\end{figure}

Figure~\ref{nu_fig3} shows in more detail again the inverse slope
parameter as a function of particle mass for central Pb+Pb
collisions at 158 GeV with data from NA44~\cite{na44flow},
NA49~\cite{NA49_QM97}, and WA97~\cite{WA97_QM97}. The linear
increase from pions to deuterons is indicated by the solid line.
The inverse slope parameters for multi-strange baryons and for
$\phi$  mesons clearly deviate from this general trend. Those
particles do not acquire the flow velocity appropriate to their
mass during the expansion. They interact weakly with the expanding
system since their hadronic cross sections are reduced compared to
non-strange hadrons~\cite{multi-strange}. The observed trend is
confirmed by RQMD calculations~\cite{rqmd}, also shown in
Figure~\ref{nu_fig3}. In RQMD, the interactions are modelled by
hadronic scattering alone. The consistency between the model and
experimental data clearly indicates that this trend is due to
scattering in the hadronic phase.

For light hadrons, the bigger their mass the higher the value of
the inverse slope parameter.  The multi-strange hadrons like Xi
and Omega, on the other hand, do not follow this trend. Their
inverse slope appears to decrease. As discussed in
Reference~\cite{multi-strange}, this different behavior may be
caused by the small hadronic cross sections of the multi-strange
hadrons. They decouple early from the system and do not acquire
the degree of collectivity characteristic for the abundantly
produced light hadrons. As we will see later, the situation will
change when the beam energy is increased at RHIC.

\subsection{Spectra and Radial Flow at RHIC}

\begin{figure}
\centerline{\psfig{file=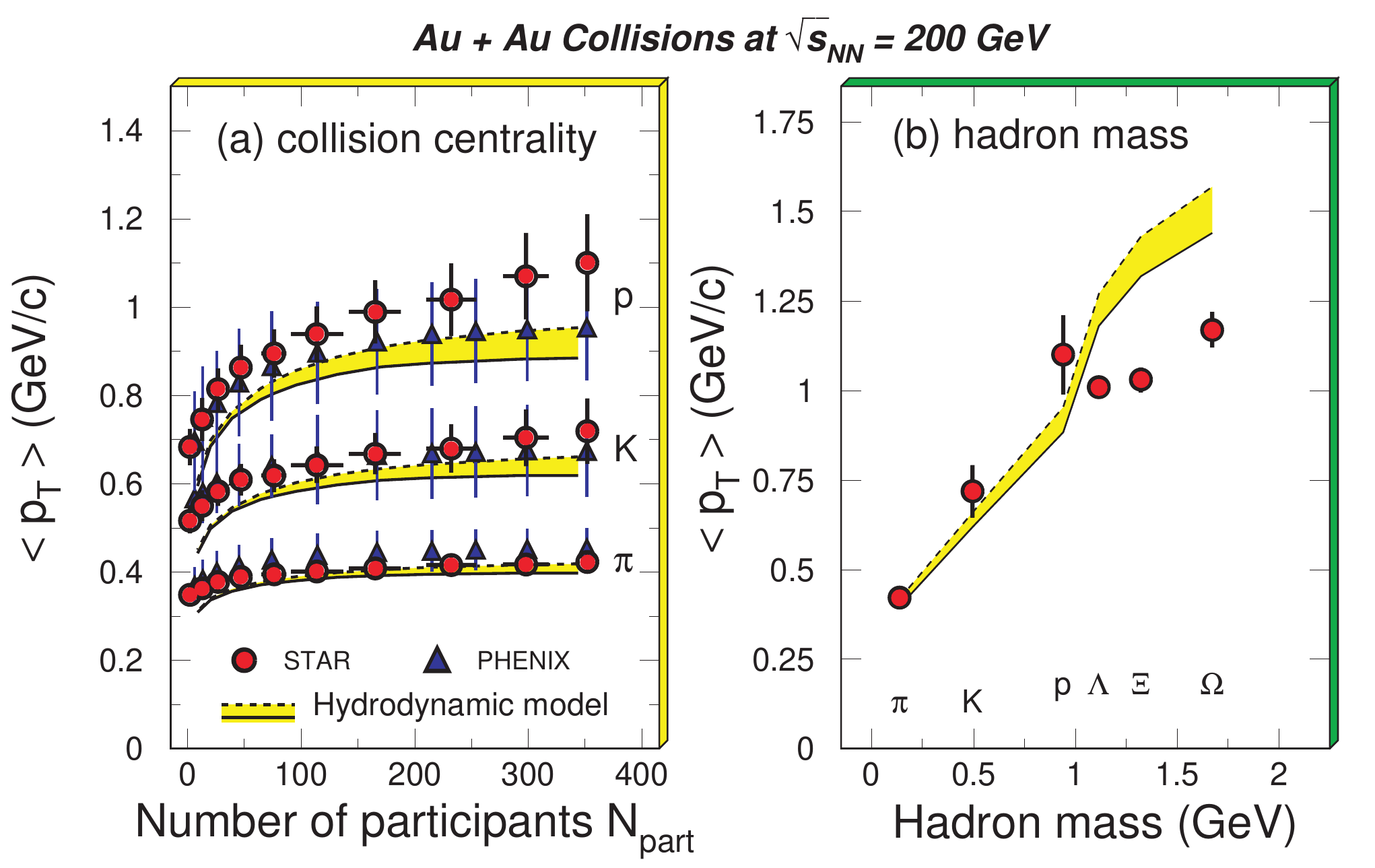,height=10.0cm}}
\caption{Average transverse momenta for pions, kaons, and protons
as a function of collision centrality (left panel) and  from
central collisions as a function of the hadron mass (right panel).
Hydrodynamic model results~\cite{Kolbx} as shown as bands.}
\label{nu_fig8}
\end{figure}

In the previous Section, we compared the systematic results from
AGS and SPS with predictions from transport models. In
case of the transport approach, the driving force is rescattering.
No assumption is made regarding the equation of state and the
nature of thermalization. A sufficient number of rescatterings
will naturally lead to collectivity and eventually to local
equilibrium. We can also look at the collisions from the point of
view of hydrodynamics. In Figure~\ref{nu_fig8} we compare the data
with hydrodynamic model calculations~\cite{Kolbx}. In the
following, we will assume that we can identify bulk properties by
measuring the most abundantly produced particles, like pions,
kaons, and protons.

The values of the mean transverse momentum of bulk particles, from
PHENIX~\cite{PHENIX_white} (triangles) and STAR~\cite{STAR_white}
are shown in the left panel of Figure~\ref{nu_fig8} as a function
of collision centrality. Also shown in the plot are the
hydrodynamic model calculations where the bands indicate the
uncertainties caused by the initial conditions.  Within error
bars, the data for bulk particles are well reproduced by the model
calculations, implying the validity of the application of
hydrodynamic approach for bulk productions in high-energy nuclear
collisions.  In the right panel of Figure~\ref{nu_fig8}  we
compare the $<p_T>$  of hadrons as a function of hadron mass for
pions, kaons, protons, $\Lambda$, $\Xi$, and $\Omega$ from central
Au+Au collisions. In the mass region below 1 GeV the hydrodynamic
calculations can account for the measured results reasonably well.
In the mass region above 1 GeV the measured mean transverse
momenta are below model predictions. In this region most of the
produced hadrons contain one or more strange quarks.   The
interaction cross sections of strange (or multi-strange) hadrons
are smaller than those of non-strange
hadrons~\cite{multi-strange}. As a result, they decouple from the
system relatively early. Therefore the values of the $<p_T>$ are
lower than those from hydrodynamic model calculations which can
reproduce the mean transverse momentum for bulk particles, but
over-predict  $<p_T>$ for multi-strange hadrons.

\begin{figure}
\centerline{\psfig{file=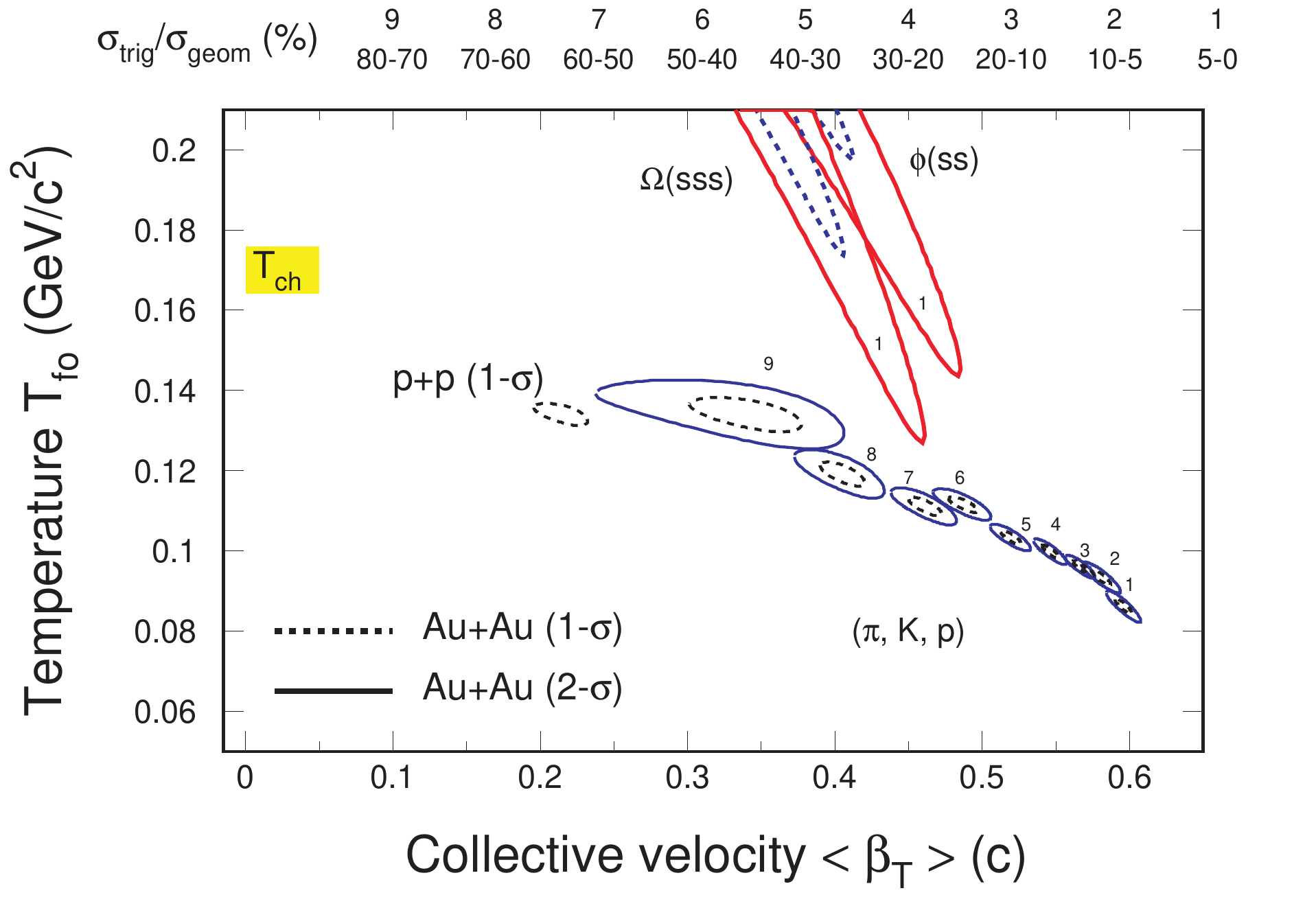,height=11.0cm}}
\caption{ $\chi^2$ contours from blast wave  fits for $\pi$, K,
and protons and for the multi-strange hadrons, $\phi$ and
$\Omega$. Numbers on top indicate the centrality selection. Dashed
and solid lines represent 1-$\sigma$ and 2-$\sigma$ contours,
respectively.} \label{nu_fig9}
\end{figure}

Figure~\ref{nu_fig9} shows $\chi^2$ contours in the temperature
versus collective velocity plane. Dashed and solid lines represent
1-$\sigma$ and 2-$\sigma$ contours, respectively,  extracted from
blast wave fits \cite{heinz93}.  Numbers on the top indicate the
centrality selection. For bulk particles 9 centrality bins are
presented.  For $\phi$ and $\Omega$ only the results from the most
central collisions are shown. Contours from p+p
collisions~\cite{pikp03} are also shown.

With increasing centrality the bulk temperature decreases and the
collective velocity increases. The velocity becomes as high as
60\% of the speed of light for the most central collisions. In
contrast, the multi-strange hadrons, $\phi$ and $\Omega$, emitted
in central collisions have a temperature of T $\sim$ 180 MeV and
an average velocity of $\beta \sim 0.4$. From Figure~\ref{nu_fig7}
we also know that their temperature and expansion velocity is not
sensitive to the collision centrality.

The thermal freeze-out temperature for the bulk is about 100 MeV.
For the multi-strange hadrons it is about 170 MeV, the same as the
bulk chemical freeze-out temperature
\cite{nxu01,pbm9596,becattini98,cleymans98,letessier94}. This
temperature is close to the value of the expected phase transition
temperature~\cite{karsch}. Again, we can conclude that the
multi-strange hadrons do not participate in the full development
of the expansion, they decouple from the system near the
hadronization point, $T \sim 170$ MeV and $\beta \sim 0.4$. A
mechanism that could lead to early freeze-out is the effect that
the hadronic cross sections for strange and especially
multi-strange hadrons are reduced~\cite{multi-strange}, as
explained earlier.

An alternative scenario has been proposed by Broniowski and
Florkowski~\cite{bf}. They assume a single freeze-out and explain
the apparently low freeze-out temperature and the large collective
velocity for the bulk as an effect of resonance decay. However, it
is not clear that this model can also explain the  data on
resonance production~\cite{zxu03}.

\begin{figure}
\centerline{\psfig{file=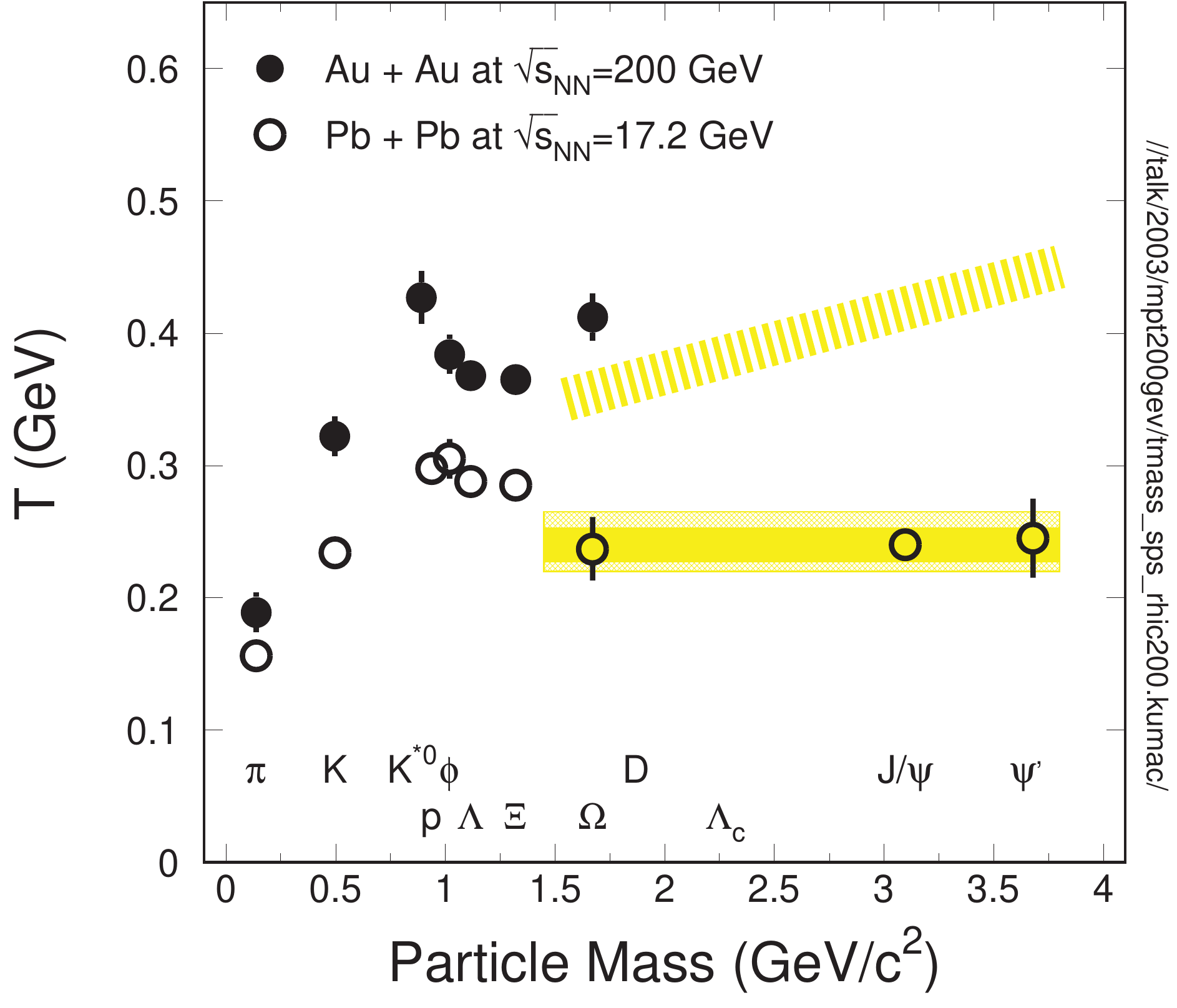,height=7.5cm}}
\caption{Inverse slope parameter,  $T$,  as a
  function of hadron mass for Pb+Pb central collisions at SPS
  (open circles) and Au+Au central collisions at RHIC
  (filled circles). The hatched area indicates the expectations discussed in the text.}
\label{nu_fig10}
\end{figure}


Figure~\ref{nu_fig10} shows the inverse slope parameters of
identified hadrons as a function of particle mass. At SPS energies
the multi-strange and charm hadrons do not show an increase as
mass increases. This is indicated by the solid band. At RHIC
energy, however, the inverse slope parameters for $\Xi$ and
$\Omega$ seem to increase with mass, indicating that the
collectivity  developed already in the partonic
phase~\cite{nxu01}. The hatched band indicates where future
data points would have to fall for this assumption to be correct.


\subsection{Partonic Collectivity at RHIC}

More conclusive evidence for flow at the parton level can be derived from elliptic flow measurements. Elliptic flow is described in a separate article of this Volume~\cite{Posk_08}. We will only present a limited set of data here.

\begin{figure}
\centerline{\psfig{file=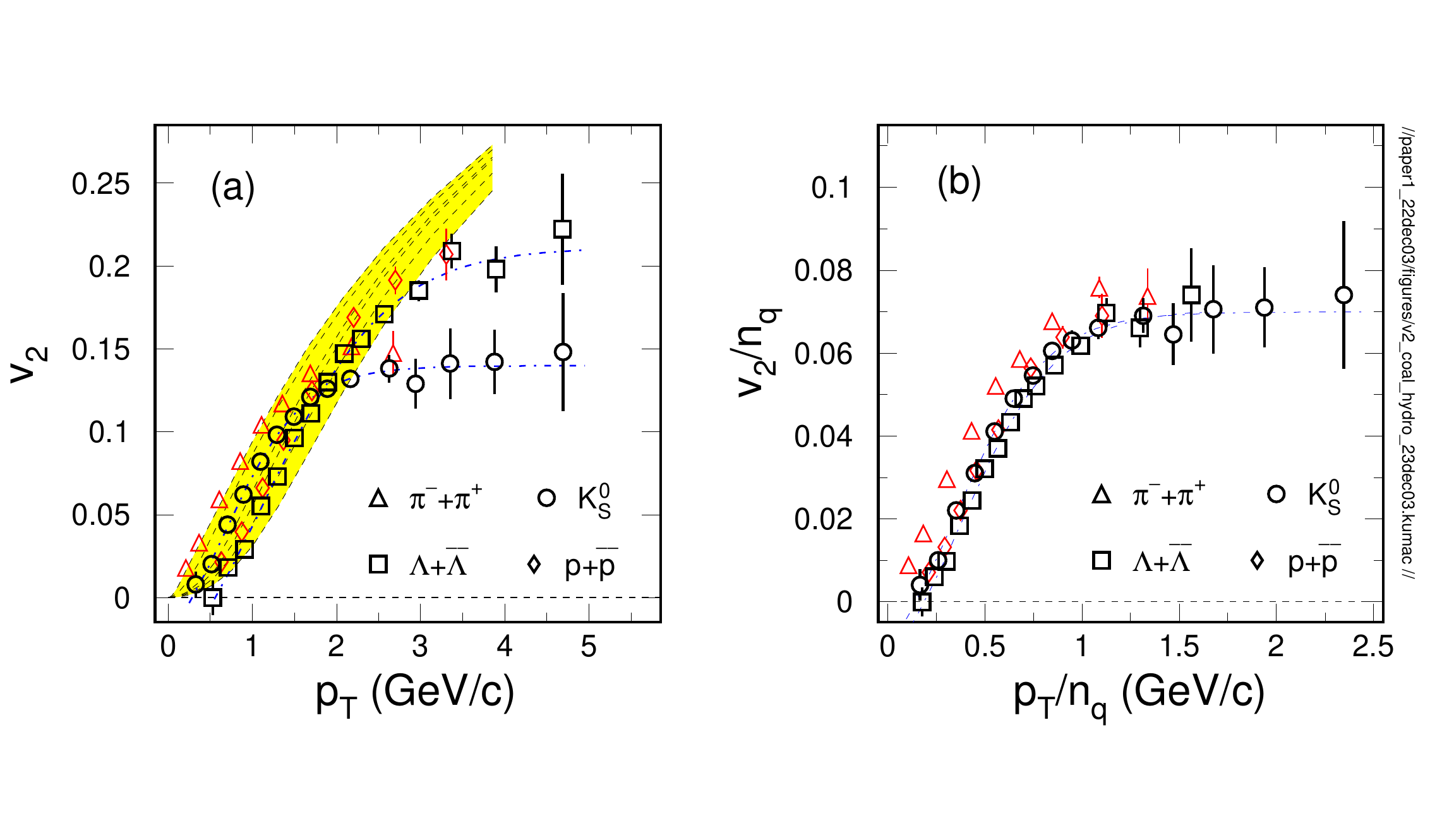,width=\textwidth}}
\caption{Left panel:  The $v_2$($p_T$) for $\pi$, K$_0^S$, p,
  and $\Lambda$ from minimum bias \auau collisions at \rts = 200 GeV
  \cite{phenixv2}. Right panel:
  $v_2$ and $p_T$ scaled by the number of constituent quarks. The dashed lines
  represent the elliptic flow of $\pi$, K, p, $\Lambda$, $\Xi$, and
 $ \Omega$ from hydrodynamic calculations. Dot-dashed lines
  are the fits to K$_0^S$ and $\Lambda$ $v_2$ distributions. }
\label{nu_fig11}
\end{figure}

The left panel of Figure~\ref{nu_fig11}  shows the measured
elliptic flow, $v_2$, from \auau collisions at \rts = 200 GeV for
$\pi$, K$_0^S$, $p$, and $\Lambda$  from PHENIX~ \cite{phenixv2} and
STAR~\cite{STAR_white}. From top to bottom the dashed lines
represent the elliptic flow of $\pi$, K, p, $\Lambda$, $\Xi$, and
$\Omega$ from hydrodynamic calculations \cite{pasi01}. At low
$p_T$, the  hydrodynamic calculations can well reproduce the \v2
measurements. At higher $p_T$,  \v2 becomes saturated and
hydrodynamic results over-predict the data. The baryons saturate
above 3 GeV/c with \v2 $\sim$ 0.2. The mesons saturate  at lower
values of $v_2$. When scaling the values of $v_2$ and $p_T$ with
the number of constituent quarks (NCQ) of the corresponding
hadrons, all particles should fall on a single curve, as
demonstrated in the right panel of  Figure~\ref{nu_fig11}. This
was predicted by coalescence models~\cite{coal}.

Scaling with the number of constituent quarks or with the number
of valence quarks has been taken as strong empirical evidence
that the constituents that acquire collectivity in the expansion
are partons (quarks) and not hadrons.


The $v_2$ values in Figure~\ref{nu_fig11} are fitted with an
empirical functional form. The fit results for K$_S^0$ and
$\Lambda$ are shown as dot-dashed lines in both panels of
Figure~\ref{nu_fig11}.  For kaon, proton, and lambda the scaling
works well for small transverse momenta  ($p_T/n_q \le 2.5$
GeV/c). Pions  do not follow the scaling. A large
fraction of pions from heavy systems is not produced directly, but
comes from the decay of resonances~ \cite{zxu03,coal2}. At
mid-rapidity as much  as  80\% of pions are from resonance decays.
The dominant sources for pion production are $\rho$, K$^*$,
K$_S^0$ and baryon resonances like $\Lambda$.  When taking
resonance decays into account, the pion $v_2$ follows the scaling
with the number of constituent quarks~\cite{Dong_04}.

The scaling of the transverse momentum dependence of $v_2$ with
the number of constituent quarks of the produced hadrons is a
powerful indication that the constituents that acquire flow are
not the hadrons, but partons at an early stage of the collision.
This is the strongest indication so far for deconfinement and a
hint for the existence of a Quark-Gluon Plasma.

\begin{figure}
\centerline{\psfig{file=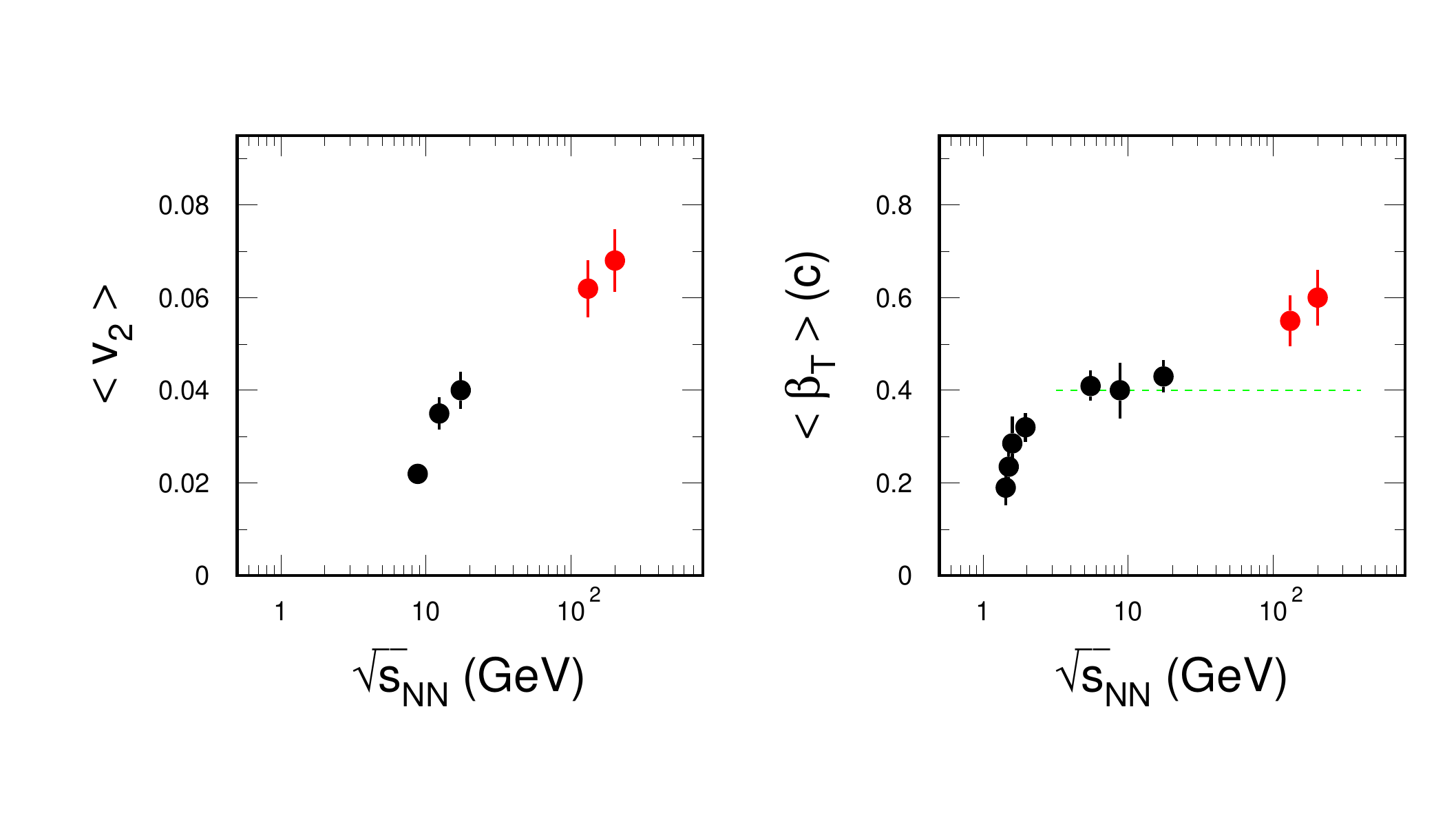,width=16cm}}
\caption{Energy dependence of the
  elliptic flow $v_2$ from minimum biased \auau or \pbpb
  interactions (left panel) and of  the radial flow velocity parameter
  $\langle \beta_T \rangle$ from central \auau or \pbpb
  collisions (right panel).}
\label{nu_fig13}
\end{figure}

Let us close this section by showing the energy dependence of the
collective flow observables in Figure~\ref{nu_fig13}. The elliptic
flow $v_2$, and the radial flow parameters, $\beta_T$, increase as
a function of collision energy. Above 6 GeV $v_2$ increases
monotonically implying that more and more early flow develops as
the collision energy increases. Assuming that hadronic
rescattering effects have already been maximized at SPS energies
(\rts $\approx$ 17 GeV), the net increase in the transverse
velocity at RHIC should be  due to partonic interactions.

\subsection{Outlook}

In the near future, it is important to quantify the partonic
collective flow with high statistics data of $v_2$ measurements
for all hadrons. Elliptic flow of $\phi$ mesons will be  important
because $\phi$ mesons are produced through quark coalescence and
not through the kaon fusion channel~\cite{kaon_ref}. These results
will provide direct information on the partonic phase.

Gauging thermalization is an important next step in the quest for
the equation of state and for the existence of a Quark-Gluon
Plasma. In lattice QCD the QGP temperature at top RHIC energy is
of the order of 0.3 to 0.5 GeV~\cite{shuryak78,rapp00}. The mass
of a charm quark is much heavier than this temperature. This means
thermal production of charm quarks is negligible. Charm quarks are
produced in first chance collisions. Thermalization of the light
quarks might be tested by measuring the elliptic flow of heavy
quarks. Heavy quarks can develop flow only if they are dragged
along during the expansion by frequent collisions with light
quarks~\cite{parton-wind}. Frequent interactions amongst light
quarks will lead to thermalization. The measurement of a finite
$v_2$ value for electrons from the semi-leptonic decay of
heavy-flavor mesons at high transverse momentum~\cite{phenix_v2}
cannot yet be taken as indication of charm quark flow in the
hydrodynamic sense.

It is also very important to pursue a temperature measurement with
thermal photons and di-leptons. NA60~\cite{NA60} recently obtained
"temperatures" for di-muons from the continuum that fit well in
the mass systematics of Figure~\ref{nu_fig10}. Thermalization of
matter is a necessary condition for measuring the temperature of
the QGP. The temperature is yet another unknown parameter in the
determination of the equation of state. Thermalization is an
important condition for our ability to  explore the QCD phase
diagram, for example by searching for the tricritical
point~\cite{tricritical} in the upcoming energy scan program at
RHIC and later at FAIR.

\section{Summary}

In heavy ion collisions at energies of 1 to 2 \AGeV densities of
two to three times normal nuclear matter density are reached.  \kp
mesons production is of special
 interest as \kp are produced close to the NN threshold.
Their production yield is very sensitive to the energy balance in
the fireball and thus to the properties of the nuclear equation of
state.  \kp interact only elastically and their yield carries
information from the early, high-density stage of the collision.
From the comparison of the measured yields with model calculations
the stiffness of nuclear matter has been extracted. Pions and
K$^-$ interact strongly with the medium. They are emitted late in
the expansion phase.

Kaon emission close to the threshold is a sensitive tool to
identify effects of in-medium KN potentials. Potentials might
cause a change of the effective thresholds and of the yields.
Presently, experimental data and the understanding of the model
parameters are not good enough to distinguish the effects of
potentials from other effects, like scattering and absorption.
 \kp and K$^0$
spectra at very low momenta are expected to be more sensitive.
With the upcoming FAIR facility at GSI it will be possible to
study potential effects through high statistics measurements by
including also charm production.

At the higher energies strong, pressure driven collective effects
have been observed. The fact that elliptic flow scales with the
number of constituent quarks and that the multi-strange hadrons
decouple from the system early have been taken as strong
indication that collectivity develops in the partonic stage and
that hadronic interactions play a minimal role for the development
of elliptic flow. We can conclude that a partonic state has been
created.

Deconfinement is not sufficient for the existence of a QGP.
Equally important is the concept of thermalization. The heavy
flavor program at RHIC and at the LHC will shed new light on this
important question.

\section*{Acknowledgments}

We would like to thank  J. Aichelin, M. Bleicher, P.
Braun-Munzinger, J. Cleymans, S. Esumi, A. F\"orster, C. Hartnack,
B. Jacak, U. Heinz, I. Kraus, Y. Leifels, C. M\"untz, A.
Poskanzer, K. Redlich, B. Schlei, E. Shuryak, H. St\"ocker, J.
Sullivan, S. Voloshin, and H. van Hecke for stimulating
discussions and suggestions. This work was supported in part by
the U.S. Department of Energy
 under Contract No. DE-AC03-76SF00098 and by the Bundesministerium f\"ur Bildung und Forschung (BMBF).


\bibliography{review}

\begin{thebibliography}{10}

\bibitem{satzqm02}
H. Satz, Nucl. Phys. {\bf A715}, 3c (2003).

\bibitem{karsch}
F. Karsch, Nucl. Phys. {\bf A698}, 199c (1996).

\bibitem{STAR_white}
J. Adams, {\it et al.}, STAR Collaboration, Nucl. Phys. {\bf A757}, 102 (2005).

\bibitem{Nagamiya_82a}
S. Nagamiya, Phys. Rev. Lett. {\bf 49}, 1383 (1982).

\bibitem{PDG}
Particle Data Group, Phys. Rev. D {\bf 54}, 1 (1996).

\bibitem{Kaplan}
D.B.~Kaplan and A.E.~Nelson, Phys. Lett. {\bf B175}, 57 (1986).

\bibitem{Schaffner}
J.~Schaffner, J.~Bondorf, and I.N.~Mishustin, Nucl. Phys. {\bf A625}, 325
  (1997).

\bibitem{Aich_Rev}
J. Aichelin and J. Schaffner-Bielich, this Volume.

\bibitem{iqmd_PR}
C.~Hartnack, Y. Leifels, H. Oeschler, and J. Aichelin, Physics Report, in
  preparation.

\bibitem{Nagamiya_81}
S. Nagamiya, {\it et al.}, Phys. Rev. C {\bf 24}, 971 (1981).

\bibitem{Nagamiya_82}
S. Nagamiya, {\it et al.}, Phys. Rev. Lett. {\bf 48}, 1780 (1982).

\bibitem{Wolf_82}
K.L. Wolf, {\it et al.}, Phys. Rev. C {\bf 26}, 2572 (1982).

\bibitem{Brockmann_84}
R. Brockmann, {\it et al.}, Phys. Rev. Lett. {\bf 53}, 2012 (1984).

\bibitem{Hayashi_88}
S. Hayashi, {\it et al.}, Phys. Rev. C {\bf 38}, 1229 (1988).

\bibitem{Siemens_79}
P.J. Siemens and J.O. Rasmussen, Phys. Rev. Lett. {\bf 42}, 880 (1979).

\bibitem{Harris_87}
J.W. Harris, {\it et al.}, Phys. Rev. Lett. {\bf 58}, 463 (1987).

\bibitem{Cugnon_81}
J. Cugnon, T. Mizutani, and J. Vandermeulen, Nucl. Phys. {\bf A352}, 505
  (1981).

\bibitem{Harris_85}
J.W. Harris, {\it et al.}, Phys. Lett. {\bf 153B}, 377 (1985).

\bibitem{Schnetzer_82}
S. Schnetzer, {\it et al.}, Phys. Rev. Lett. {\bf 49}, 989 (1982).

\bibitem{Schnetzer_89}
S. Schnetzer, {\it et al.}, Phys. Rev. C {\bf 40}, 640 (1989); Erratum-ibid.
  {\bf 41}, 1320 (1990).

\bibitem{Shor_89}
A. Shor, {\it et al.}, Phys. Rev. Lett. {\bf 63}, 2192 (1989).

\bibitem{Shor_82}
A. Shor, {\it et al.}, Phys. Rev. Lett. {\bf 48}, 1597 (1982).

\bibitem{Harris_81}
J.W. Harris, {\it et al.}, Phys. Rev. Lett. {\bf 47}, 229 (1981).

\bibitem{Carroll_89}
J.B. Carroll, {\it et al.}, Phys. Rev. Lett. {\bf 62}, 1829 (1989).

\bibitem{Stock_86}
R. Stock, Phys. Rept. {\bf 135}, 259 (1986).

\bibitem{Senger_93}
P. Senger, {\it et al.}, KaoS Collaboration, Nucl. Instr. and Meth. {\bf A327},
  393 (1993).

\bibitem{Gobbi_93}
A. Gobbi, {\it et al.}, FOPI Collaboration, Nucl. Instr. and Meth. {\bf A324},
  156 (1993).

\bibitem{reisdorf-pionen}
W. Reisdorf, {\it et al.}, FOPI Collaboration, Nucl.Phys. {\bf A781}, 459
  (2007).

\bibitem{KaoS}
A. {F\"orster}, {\it et al.}, KaoS Collaboration, Phys.~Rev.~C {\bf 75}, 024906
  (2007).

\bibitem{Randrup_80}
J. Randrup and C.M. Ko, Nucl. Phys. {\bf A343}, 519 (1980); Erratum-ibid. {\bf
  A411}, 537 (1983).

\bibitem{iqmd}
C.~Hartnack, {\it et al.}, Eur. Phys. J. {\bf A1}, 151 (1998).

\bibitem{Fuchs_Rev_06}
C.~Fuchs, Prog. Part. Nucl. Phys. {\bf 56}, 1 (2006).

\bibitem{hong_fopi}
B. Hong, {\it et al.}, FOPI Collaboration, Phys. Lett. {\bf B407}, 115 (1997).

\bibitem{sturm}
C. Sturm, {\it et al.}, KaoS Collaboration, Phys. Rev. Lett. {\bf 86}, 39
  (2001).

\bibitem{Ko_84}
C.M.~Ko, Phys. Lett. {\bf B138}, 361 (1984).

\bibitem{Foerster_03}
A.~{F\"ors}ter, {\it et al.}, KaoS Collaboration, Phys. Rev. Lett. {\bf 91},
  152301 (2003).

\bibitem{Oeschler_00}
H.~Oeschler, J.~Phys. G {\bf 27}, 257 (2001).

\bibitem{Hartnack_03}
C. Hartnack, H. Oeschler, and J. Aichelin, Phys. Rev. Lett. {\bf 90}, 102302
  (2003); Erratum-ibid. {\bf 93}, 149903 (2004).

\bibitem{STARratio}
B.I. Abelev, {\it et al.}, Star Collaboration, e-Print:arXiv0808.2041.

\bibitem{Cleymans_99}
J.~Cleymans, H.~Oeschler, and K.~Redlich, Phys. Rev. C {\bf 59}, 1663 (1999);
  Phys. Lett. {\bf B485}, 27 (2000).

\bibitem{STARpaper}
J. Adams, {\it et al.}, STAR Collaboration, Phys. Lett. {\bf B567}, 167 (2003).

\bibitem{Averbeck_03}
R. Averbeck, {\it et al.}, TAPS Collaboration, Phys. Rev. C {\bf 67}, 024903
  (2003).

\bibitem{Sturm_diss}
C. Sturm, Ph.D. thesis, TU Darmstadt 2001.

\bibitem{Schwalb_94}
O. Schwalb, {\it et al.}, Phys.~Lett.~ {\bf B321}, 20 (1994).

\bibitem{Muentz_95}
C. {M\"untz}, {\it et al.}, KaoS Collaboration, Z.~Phys.~{\bf A352}, 175
  (1995).

\bibitem{Cleymans_04}
J.~Cleymans, A.~{F\"or}ster, H.~Oeschler, K.~Redlich, and F.~Uhlig, Phys. Lett.
  {\bf B603}, 146 (2004).

\bibitem{Wagner_98}
A. Wagner, {\it et al.}, KaoS Collaboration, Phys. Lett. {\bf B420}, 20 (1998);
  Phys. Rev. Lett. {\bf 85}, 18 (2000); A. Wagner, Ph.D. thesis, TU Darmstadt
  1996.

\bibitem{Knoll_08}
J. Knoll, e-Print: arXiv:0803.2343.

\bibitem{Weinhold_98}
W. Weinhold, B. Friman, and W. {N\"or}enberg, Phys. Lett. {\bf B433}, 236
  (1998).

\bibitem{Sullivan_82}
J.P. Sullivan {\it et al.}, Phys. Rev. C {\bf 25}, 1499 (1982).

\bibitem{Laue}
F. Laue, {\it et al.}, KaoS Collaboration, Phys. Rev. Lett. {\bf 82}, 1640
  (1999).

\bibitem{Menzel}
M. Menzel, {\it et al.}, KaoS Collaboration, Phys. Lett. {\bf B495}, 26 (2000).

\bibitem{Wisn}
K. Wisniewski, {\it et al.}, FOPI Collaboration, Eur. Phys. J. {\bf A9}, 515
  (2000).

\bibitem{Merschmeyer_07}
M. Merschmeyer, {\it et al.}, FOPI Collaboration, Phys. Rev. C {\bf 76}, 024906
  (2007).

\bibitem{HADES_AS}
G. Agakishiev, {\it et al.}, HADES Collaboration, arXiv:0902.3487.

\bibitem{fuchs}
C.~Fuchs, A.~Faessler, E.~Zabrodin, and Y.~M.~Zheng, Phys. Rev. Lett. {\bf 86},
  1974 (2001).

\bibitem{iqmd_eos}
Ch. Hartnack, H. Oeschler, and J. Aichelin, Phys. Rev. Lett. {\bf 96}, 012302
  (2006).

\bibitem{ritter97}
W. Reisdorf and H.G. Ritter, Ann. Rev. Nucl. Part. Sci. {\bf 47}, 663 (1997).

\bibitem{Shin}
Y. Shin, {\it et al.}, KaoS Collaboration, Phys. Rev. Lett. {\bf 81}, 1576
  (1998).

\bibitem{Uhlig}
F. Uhlig, {\it et al.}, KaoS Collaboration, Phys. Rev. Lett. {\bf 95}, 012301
  (2005).

\bibitem{starp130}
J. Adams, {\it et al.}, STAR Collaboration, Phys. Rev. C {\bf 70}, 041901
  (2004).

\bibitem{na44flow}
I.G. Bearden, {\it et al.}, NA44 Collaboration, Phys. Rev. Lett {\bf 78}, 2080
  (1997); M. Kaneta, Ph.D Thesis, Hiroshima University (1999).

\bibitem{heinz93}
E. Schnedermann, J. Sollfrank, and U. Heinz, Phys. Rev. C {\bf 48}, 2462
  (1993).

\bibitem{starpion}
M.~Calder\'{o}n~de~la~Barca~S\'{a}nchez, {\it et al.}, STAR Collaboration,
  Nucl. Phys. {\bf A698}, 503c (2002).

\bibitem{starkaon}
C. Adler, {\it et al.}, STAR Collaboration, Phys. Lett. {\bf B595}, 143 (2004).

\bibitem{fig22_ref}
J. Adams {\it et al.}, STAR Collaboration, Phys. Rev. Lett. {\bf 98}, 060301
  (2007); B.I. Abelev, {\it et al.}, STAR Collaboration, Phys. Rev. Lett. {\bf
  99}, 112301 (2007).

\bibitem{ua196}
G. Bocquet, {\it et al.}, UA1 Collaboration, Phys. Lett. {\bf B366}, 441
  (1996).

\bibitem{E802}
L. Ahle, {\it et al.}, E802 Collaboration, Phys. Rev. C {\bf 60}, 064901
  (1999); Phys. Rev. C {\bf 57}, 466 (1998).

\bibitem{xu_NA44}
N. Xu, {\it et al.}, NA44 Collaboration, Nucl. Phys. {\bf A610}, 175c (1996).

\bibitem{nxu01}
N. Xu and M. Kaneta, Nucl. Phys. {\bf A698}, 306c (2001).

\bibitem{NA49_QM97}
G. Roland, {\it et al.}, NA49 Collaboration, Nucl. Phys. {\bf A638}, 91c
  (1998).

\bibitem{WA97_QM97}
I. Kralik, {\it et al.}, WA97 Collaboration, Nucl. Phys. {\bf A638}, 115c
  (1998).

\bibitem{multi-strange}
H. van Hecke, H. Sorge, and N. Xu, Phys. Rev. Lett. {\bf 81}, 5764 (1998); Y.
  Cheng, F. Liu, Z. Liu, K. Schweda, and N. Xu, Phys. Rev. C {\bf 68}, 034901
  (2003).

\bibitem{rqmd}
H. Sorge, Phys. Rev. C \textbf{52}, 3291 (1992).

\bibitem{Kolbx}
P. Kolb and R. Rapp, Phys. Rev. C {\bf 67}, 044903 (2003).

\bibitem{PHENIX_white}
S.S. Adler, {\it et al.}, PHENIX Collaboration, Nucl. Phys. {\bf A757}, 184
  (2005).

\bibitem{pikp03}
J. Adams, {\it et al.}, STAR Collaboration, Phys. Lett. {\bf B616}, 8 (2005).

\bibitem{pbm9596}
P. Braun-Munzinger, J. Stachel, J. Wessels, and N. Xu, Phys. Lett. {\bf B344},
  43 (1995); P. Braun-Munzinger, I. Heppe, and J. Stachel, Phys. Lett. {\bf
  B465}, 15 (1999).

\bibitem{becattini98}
F. Becattini, M. Gazdzicki, and J. Sollfrank, Eur. Phys. J. {\bf C5}, 143
  (1998); F. Becattini, Z. Phys. {\bf C69}, 485 (1996); F. Becattini and U.
  Heinz, Z. Phys. {\bf C76}, 269 (1997).

\bibitem{cleymans98}
J. Cleymans and K. Redlich, Phys. Rev. Lett. {\bf 81}, 5284 (1998).

\bibitem{letessier94}
J. Letessier, J. Rafelski, and A. Tounsi, Phys. Lett. {\bf B328}, 499 (1994);
  M. Kaneta, {\it et al.}, NA44 Collaboration, J. Phys. G {\bf 23}, 1865
  (1997).

\bibitem{bf}
W. Broniowski and W. Florkowski, Phys. Rev. C {\bf 65}, 024905 (2002); Phys.
  Rev. Lett. {\bf 87}, 272302 (2001).

\bibitem{zxu03}
Z. Xu, J. Phys. G {\bf 30}, S325 (2004).

\bibitem{Posk_08}
S. Voloshin, A.M. Poskanzer, and R. Snellings, this Volume.

\bibitem{phenixv2}
S.S. Adler, {\it et al.}, PHENIX Collaboration, Phys. Rev. Lett. {\bf 91},
  182301 (2003).

\bibitem{pasi01}
P. Huovinen, P. Kolb, U. Heinz, P.V. Ruuskanen, and S. Voloshin, Phys. Lett.
  {\bf B503}, 58 (2001).

\bibitem{coal}
Z. Lin and C. Ko, Phys. Rev. Lett. {\bf 89}, 202302 (2002); R.J. Fries, B.
  Mueller, C. Nonaka, and S.A. Bass, Phys. Rev. Lett. {\bf 90}, 202303 (2003);
  D. Molnar and S. Voloshin, Phys. Rev. Lett. {\bf 91}, 092301 (2003).

\bibitem{coal2}
R.J. Fries, B. Mueller, C. Nonaka, and S.A. Bass, Phys. Rev. C {\bf 68}, 044902
  (2003); C. Nonaka, B. Mueller, M. Asakawa, S.A. Bass, and R.J. Fries, Phys.
  Rev. C {\bf 69}, 031902 (2004).

\bibitem{Dong_04}
X. Dong, S. Esumi, P. Sorensen, N. Xu, and Z. Xu, Phys. Lett. {\bf B597}, 328
  (2004).

\bibitem{kaon_ref}
J. Adams, {\it et al.}, STAR Collaboration, Phys. Lett. {\bf B612}, 181 (2005)
  and references therein.

\bibitem{shuryak78}
E.V. Shuryak, Phys. Lett. {\bf B78}, 150 (1978).

\bibitem{rapp00}
R. Rapp, Phys. Rev. C {\bf 63}, 054907 (2001).

\bibitem{parton-wind}
X. Zhu, N. Xu, and P. Zhuang, Phys. Rev. Lett. {\bf 100}, 152301 (2008).

\bibitem{phenix_v2}
A. Adare, {\it et al.}, PHENIX Collaboration, Phys. Rev. Lett. {\bf 98}, 172301
  (2007).

\bibitem{NA60}
R. Arnaldi, {\it et al.}, NA60 Collaboration, Phys. Rev. Lett. {\bf 100},
  022302 (2007).

\bibitem{tricritical}
M.A. Stephanov, K. Rajagopal, and E.V. Shuryak, Phys. Rev. Lett. {\bf 81}, 4816
  (1998).

\end{thebibliography}
\bibliographystyle{unsrt}

\end{document}